\DocumentMetadata{lang=en-GB,tagging=on}
\documentclass[a4paper]{article}
\usepackage[unicode,breaklinks=true,final]{hyperref}
\usepackage[utf8]{inputenc}
\usepackage[T1,T2A]{fontenc}
\usepackage[unicode,final]{hyperref}
\usepackage{amsmath,amssymb,amsthm,orcidlink,cleveref,mleftright,mathtools,booktabs}
\usepackage[noadjust,nosort]{cite}
\usepackage{CJKutf8}
\usepackage[final]{microtype}
\usepackage[latin,russian,french,ngerman,greek,british]{babel}
\usepackage{CJKutf8}
\DeclareMathOperator\tr{tr}
\DeclareMathOperator\csch{csch}

\usepackage{tikz}
\usetikzlibrary{decorations.pathreplacing}
\newcommand{\gd}{g_\mathrm{D1NC}}
\newcommand{\gsnc}{g_\mathrm{SNC}}
\usepackage[osf]{newpxtext}
\usepackage[varbb,slantedGreek]{newpxmath}
\renewcommand\pi\piup
\renewcommand\varepsilon\varepsilonup
\renewcommand\Delta\Deltaup

\title[pdftitle={{Non-Relativistic Limits of 𝒩 = 4 Supersymmetric Yang–Mills Theory and S‐duality}}]{Non-Relativistic Limits of \(\mathcal N=4\) Supersymmetric Yang--Mills Theory and S-duality}

\author[pdfauthor={{Hyungrok Kim, Joseph Smith}}]{
\parbox{5cm}{\centering Hyungrok Kim
(\begin{CJK*}{UTF8}{bsmi}金炯錄\end{CJK*})\footnote{Centre for Mathematics and Theoretical Physics Research, Department of Physics, Astronomy and Mathematics, University of Hertfordshire, Hatfield, Hertfordshire\ \textsc{al10~9ab}, United Kingdom}\,\,\,\textsuperscript{\orcidlink{0000-0001-7909-4510}}\\\href{mailto:h.kim2@herts.ac.uk}{\texttt{h.kim2@herts.ac.uk}}}
\and\parbox{5cm}{\centering Joseph Smith\footnote{School of Mathematics, University of Birmingham, Watson Building, Edgbaston, Birmingham \textsc{b15~2tt}, United Kingdom}\,\,\,\textsuperscript{\orcidlink{0000-0003-0114-9109}}\\\href{mailto:j.smith.24@bham.ac.uk}{\texttt{j.smith.24@bham.ac.uk}}}
}
\begin{document}
\maketitle
\begin{abstract}
We investigate non-relativistic limits of four-dimensional maximally supersymmetric Yang–Mills theory (4d MSYM) and their relation to the nonperturbative \(\operatorname{SL}(2;\mathbb Z)\) S-duality of the relativistic theory. We construct a general family of non-relativistic limits using a Type~IIB brane set-up with a D3-brane and \((p,q)\)-strings and show that the resulting theories are topological deformations of supersymmetric Galilean Yang--Mills theory or quantum mechanics on the moduli space of BPS monopoles. The deformations of the Galilean Yang--Mills theory are the familiar \(\theta\)-term and a coupling to the monopole charge, while in the moduli space theory the only deformation is a \(\theta\)-term. This family of theories fit together into a three-dimensional moduli space with nontrivial topology, on which \(\operatorname{PSL}(2;\mathbb Z)\)-valued dualities act in a richer and more complex way than in the relativistic parent theory. In the Abelian case, we establish the duality directly using the path integral, while in the non-Abelian case we support our claim by matching the one-particle spectrum as well as the Galilean spacetime symmetries and electric/magnetic invertible one-form symmetries. 
\end{abstract}

\vspace{3.0cm}
\newpage

\tableofcontents

\section{Introduction}
Four-dimensional maximally supersymmetric (\(\mathcal N=4\)) Yang--Mills theory (4d MSYM) enjoys a plethora of symmetries, such as \(\mathcal N=4\) superconformal symmetry, a dual superconformal symmetry \cite{Drummond:2008vq} and the Yangian symmetry \cite{Drummond:2009fd} (in the planar limit), and one-form electric and magnetic centre symmetries \cite{Gaiotto:2014kfa}.
It is also at the epicentre of a number of dualities: the holographic duality to Type~IIB superstring theory on \(\operatorname{AdS}_5\times\mathbb S^5\), the double copy to four-dimensional \(\mathcal N=8\) supergravity, and (most importantly for this paper) a nonperturbative \(\operatorname{SL}(2;\mathbb Z)\) Montonen--Olive S-duality \cite{Montonen:1977sn} that descends from the S-duality of Type~IIB string theory, in which the D3-brane (whose worldvolume theory's infrared limit is 4d MSYM) is self-dual.
Unlike the case of the electromagnetic duality of Maxwell theory (and more general \(p\)-form electrodynamics), which can be seen direction from path-integral manipulations \cite{Deser:1976iy}, the S-duality of 4d MSYM is nonperturbative and cannot be easily seen from the action. Furthermore, it can changes the gauge group to the Langlands-dual group and therefore has close connections to the geometric Langlands conjecture \cite{Kapustin:2006pk,Gukov:2006jk,Witten:2007td}.

It is, therefore, natural to seek limits in which the strongly-coupled dynamics of 4d MSYM simplifies
and we can examine the S-duality more directly.
A natural choice one can make to simplify a QFT is to take a non-relativistic limit in which particle number becomes conserved; however, as 4d MSYM is conformal it is not immediately clear that any such limits exist as there is no sense in which a low-energy limit decouples degrees of freedom. The solution to this is to work on the Coulomb branch of the theory; the scalar VEV defines a mass scale and allows for particle states for which we can isolate the low-energy physics. We then turn to the question of how the non-relativistic limit should be taken, as there is not a unique way to do this. This is most easily answered by viewing the set-up within string theory, where the field theory is the IR description of a D3-brane stack and the VEV corresponds to separating the branes. Particle degrees of freedom are then engineered by stretching 1-branes between two D3-branes: the most obvious choices are for the 1-branes are fundamental strings, corresponding to W-bosons in the QFT (as can be seen directly from the open string spectrum), or D1-branes, corresponding to BPS monopoles \cite{Diaconescu:1996rk}. We can then take non-relativistic limits associated with these particles by localising onto F1 or D1-brane configurations within D3-branes, which can be implemented within the QFT using the recent BPS decoupling limit methods outlined in \cite{Blair:2023noj, Lambert:2024uue, Blair:2024aqz, Lambert:2024ncn}. For fundamental strings the resulting theory is a supersymmetric version of Galilean Yang-Mills \cite{Bagchi:2015qcw, Bagchi:2022twx, Fontanella:2024rvn}, while the theory obtained by the D1-brane limit contains the Bogomolny equation as a constraint \cite{Lambert:2018lgt, Lambert:2024yjk} and reduces to supersymmetric quantum mechanics on the moduli space of BPS monopoles (as in the Manton approximation for monopole scattering \cite{Manton:1981mp}).

Within the relativistic theory, S-duality transformations map the two types of particle into each other. Since our interpretation of the non-relativistic limits is that they isolate the low-energy excitations of their respective particles it is reasonable to propose that, despite their remarkably different structure, the two theories are in reality dual to each other \cite{Lambert:2024yjk}. More generally, we could have started with a $(p,q)$-string stretched between D3-branes and taken the corresponding near-BPS limit. As these objects are transformed into each other via \(\operatorname{SL}(2;\mathbb Z)\) transformations, the natural extension of the above proposal is then that we have a family of theories that map into each other through the action of this group. From a string-theoretic perspective, it is believed that the non-relativistic string and D1-brane limits of Type~IIB string theory continue to enjoy \(\operatorname{SL}(2;\mathbb Z)\) duality \cite{Bergshoeff:2022iss,Bergshoeff:2023ogz,Ebert:2023hba, Blair:2025prd} that maps between the two; our goal in this work is to provide evidence for this claim through the study of 4d MSYM's non-relativistic limits.

\paragraph{Results}
Starting from the flat solution of  type~IIB supergravity, we apply a general \(\operatorname{SL}(2;\mathbb R)\) transformation and construct the corresponding non-relativistic limit of four-dimensional maximally supersymmetric Yang--Mills theory. This yields a \(\operatorname{SL}(2;\mathbb R)\) family of non-relativistic theories, which are deformations of the supersymmetric Galilean Yang--Mills theory (SNC limit) \cite{Bagchi:2015qcw,Bagchi:2022twx,Fontanella:2024rvn} or of the D1NC limit \cite{Lambert:2024yjk}.

The SNC theory admits \emph{two} different deformations, parameterised by \(\theta_\mathrm{SNC}\) and \(\lambda\). The former is the limit of the relativistic $\theta$-term\footnote{Which is $c$-dependent after the $\operatorname{SL}(2;\mathbb R)$ transformation is performed.}, while the latter is only realisable in the non-relativistic theory and shifts $F_{ij}$ by $\varepsilon_{ijk} D_k X$. Using a field redefinition, we show that this shift can be rewritten as a term proportional to the monopole charge and therefore does not affect the theory's dynamics. The D1NC theory admits only the $\theta$-term deformation, with coefficient \(\theta_\mathrm{D1NC}\). The \(\operatorname{PSL}(2;\mathbb Z)\) duality transformations of the relativistic theory relates the two theories in an intricate way,
\begin{subequations} 
\begin{align}
    \operatorname{D1NC}(g,\theta)&\overset{\mathsf T}\mapsto \operatorname{D1NC}(g,\theta+2\pi) \ ,\\
    \operatorname{SNC}(g,\theta,\lambda)&\overset{\mathsf T}\mapsto \operatorname{SNC}(g,\theta+2\pi,\lambda) \ , \\
    \operatorname{D1NC}(g,\theta)
    &\overset{\mathsf S } \mapsto \operatorname{SNC}(4\pi/g,0,-g^2\theta/8\pi^2),\\
    \operatorname{SNC}(g,\theta,\lambda)&\overset{\mathsf S}\mapsto\begin{cases}
    \operatorname{D1NC}(4\pi/g,-g^2 \lambda/2)&\text{if \(\theta=0\)},\\ 
    \operatorname{SNC}\mleft(g|\theta|/2\pi,-4\pi^2/\theta,\lambda + 8\pi^2 / g^2 \theta \mright)&\text{if \(\theta\ne0\)}, \end{cases}
\end{align}
\end{subequations}
where the presence of two different deformations of the SNC theory produces a richer web of dualities at the level of the action than in the relativistic case. We show this explicitly at the level of the Abelian path integral, and give evidence for it for the $\mathfrak{g} = \mathfrak{su}(2)$ non-Abelian by computing the one-particle spectrum of the family of theories constructed from $\operatorname{SL}(2;\mathbb{Z})$ transformations of the flat SNC solution. We find that the same tower of solutions is observed in all theories, and one can match their electric and magnetic charges in a manner predicted by the form of the $\operatorname{SL}(2;\mathbb{Z})$ transformation. Finally, we show that the two types of theory enjoy the same set of spacetime and one-form symmetries, namely a spacetime Galilean conformal symmetry (with the scaling symmetries broken by the W-boson/monopole mass) as well as a \(\operatorname Z(G)\)-valued electric one-form symmetry and a \(\pi_1(G)\)-valued magnetic one-form symmetry, where \(G\) is the gauge group. While the one-form symmetry groups match the relativistic $\mathcal{N}=4$ theory we find that the electric one-form symmetry currents of the non-relativistic theories are deformed in a non-trivial way, with the magnetic one-form currents remaining undeformed.

\paragraph{Related Work and Limitations}
We do not directly relate our results to non-relativistic string theory, as discussed in \cite{Bergshoeff:2022iss,Bergshoeff:2023ogz,Ebert:2023hba}, though we expect it to be possible to construct the $\operatorname{SL}(2;\mathbb{R})$ family of back-reacting D3-brane solutions of the non-relativistic string and D1-brane limits of type IIB supergravity that give the dual descriptions of the theories we consider here. We also focus solely on the bosonic sector of the theories in this work. The fermionic sector and supersymmetry of the SNC and D1NC theories are discussed in \cite{Lambert:2024yjk, Lambert:2026}, and it would be prudent to check that this structure is not ruined by the field theory's deformations.

Amongst the symmetries and dualities of maximal supersymmetric Yang--Mills theory listed above, we have discussed non-relativistic analogues of S-duality, (super)conformal symmetry, and one-form symmetries. Notably absent is the Yangian symmetry. The non-relativistic limits have trivial perturbative scattering amplitudes but develop nontrivial dynamics in the Coulomb branch \cite{Lambert:2026}. In the relativistic case, the dual superconformal symmetry survives with deformations \cite{Craig:2011ws}, and it would be interesting to pursue this, especially in relation to the integrability of non-relativistic string theory \cite{Fontanella:2026gaq}.

\paragraph{Organisation of this Paper}
This paper is organised as follows. After a brief review of non-relativistic limits and their relation to BPS states and supergravity in \cref{sec:review}, we construct a general family of non-relativistic limits of maximally supersymmetric Yang--Mills theory in \cref{sec:construction} using a Type~IIB brane set-up. We find that there are two qualitatively different cases: one that generalises the case with purely D-string charge (D1NC limit), discussed in \cref{sec:d1nc}, and one that generalises the case with purely F-string charge (SNC limit), discussed in \cref{sec:snc}. Then we match the two families of theories with respect to S-duality in \cref{sec:duality} and discuss the symmetries of both theories in \cref{sec:symmetry}.

\section{Review of Non-Relativistic Limits from BPS Configurations of Supergravity}\label{sec:review}
\begin{figure}\centering
\begin{tikzpicture}
    \fill[gray!30] (-2,-0.5) rectangle (0,0.5);
    \draw[->] (-2,0) -- (5,0) node[right] {$E$};
    \draw (0,-0.5) -- (0,0.5) node[above] {$E=Q$};
    \draw [thick, decoration={brace, mirror, raise=0.5cm}, decorate] (0,-0.1) -- (5,-0.1) node [pos=0.5, anchor=north, yshift=-0.55cm] {states of the original theory};
\end{tikzpicture}
\begin{equation*}
\hphantom{\text{near-BPS limit}}
{\Bigg\Downarrow}\text{near-BPS limit}
\end{equation*}
\begin{tikzpicture}
    \fill[gray!30] (-2,-0.5) rectangle (0,0.5);
    \fill[gray!30] (1,-0.5) rectangle (5,0.5);
    \draw[->] (-2,0) -- (5,0) node[right] {$E$};
    \draw (0,-0.5) -- (0,0.5) node[above] {$E=Q$};
    \draw [thick, decoration={brace, mirror, raise=0.5cm}, decorate] (0,-0.1) -- (1,-0.1) node [pos=0.5, anchor=north, yshift=-0.55cm] {states of the non-relativistic theory}
    node [pos=0.5, anchor=north, yshift=-1cm] {\(E-Q=\Deltaup E \ll Q\)};
\end{tikzpicture}

\caption{
    The non-relativistic limit obtained by looking at the near-BPS configurations above a given BPS bound. The grey region indicates forbidden states.
}\label{fig:bps-diagram}
\end{figure}
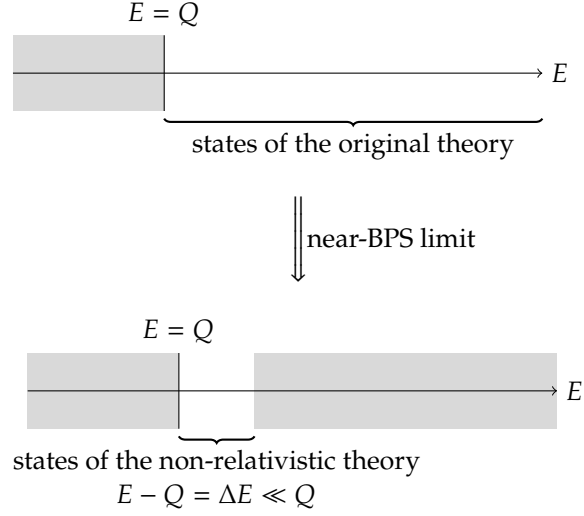

In this section, we briefly review how non-relativistic limits of brane worldvolume theories are associated to BPS configurations of the bulk supergravity theory coupling to the brane \cite{Harmark:2014mpa,Blair:2023noj,Blair:2024aqz}. The non-relativistic limit of a relativistic field theory is not in general unique: after introducing a speed of light parameter \(c\) and before taking the \(c\to\infty\) limit\footnote{Of course, since the speed of light is a dimensionful quantity what we are really doing here is introducing a dimensionless parameter $\omega$ and making the rescaling $c\to \omega c$, with $\omega$ then taken to infinity. We shall neglect the distinction here.}, one must specify how the various physical quantities (fields, coupling constants, etc.) scale with respect to \(c\). Some of these choices will be trivial and, in general, it is not a~priori obvious what the interpretation of a given \(c\to\infty\) limit means. In the presence of supersymmetry, a natural family of non-relativistic limits is obtained by considering BPS states \cite{Harmark:2014mpa}, as depicted in \cref{fig:bps-diagram}. That is, given a BPS bound \(Q\le E\) between charge \(Q\) and energy \(E\), one considers those states with charge \(Q\) and energy \(Q+\Deltaup E\) where \(\Deltaup E\ll Q\) describing small fluctuations above an extremal state saturating the BPS bound. 

This is naturally embedded in string theory when the relativistic field theory in question is the worldvolume theory of a BPS brane, where the non-relativistic limit can be interpreted as arising from near-BPS excitations of a quarter-BPS intersecting brane set-up \cite{Blair:2023noj,Lambert:2024uue, Blair:2024aqz}. Practically, one proceeds for a given brane set-up as follows. Suppose we have a $p$-brane worldvolume QFT that we would like to take a near-BPS limit of to localise onto states representing $q$-branes in the theory. To construct the limit, we must:
\begin{enumerate}
\item We couple the brane action to the relevant background supergravity fields.
\item Take the background supergravity fields to be the BPS $q$-brane solution (oriented such that the combination of the two types of branes is quarter-BPS) parameterised by a harmonic function \(H\).
\item Set \(H\) to be a power of \(c\), and send \(c\to\infty\), scaling coupling constants as necessary.
\end{enumerate}
This procedure can be used to construct a variety of non-Lorentzian limits of supersymmetric QFTs.  Such limits are believed to extend to full limits of string theory and can also be studied in the supergravity regime; the resulting theories one gets describe non-Lorentzian gravity, where the underlying geometry is a $p$-brane generalisation of Newton-Cartan (NC) geometry. We therefore refer to these as \(X\)NC limits, where $X$ denotes the brane used to take the limit (\textit{e.g.} if the limit localises us onto low-energy D$p$-brane degrees of freedom we speak of D\(p\)NC limit, if we localise onto fundamental strings we speak of an SNC limit, and so on). In particular, SNC limits \cite{Bergshoeff:2019pij, Bidussi:2021ujm, Bergshoeff:2023ogz, Harmark:2025ikv, Bergshoeff:2025uut}, D\(p\)NC limits \cite{Lambert:2024yjk,Lambert:2024ncn, Blair:2025prd}, and M2/M5 limits \cite{Blair:2021waq,Lambert:2024uue,Bergshoeff:2024nin, Lambert:2024ncn, Bergshoeff:2025grj, Blair:2025ewa} have been discussed in the literature at the level of the worldvolume QFTs and supergravity theories, to give a non-exhaustive list of references.

\paragraph{Example: Massive Scalar Particle}
We illustrate this procedure by a simple example.
Consider a free massive \(\mathcal N=1\) chiral multiplet \((\phi,\psi)\) in four dimensions. For simplicity we look at the bosonic sector only, whose action is
\begin{equation}
    S = -\int \mathrm d^4x\,\sqrt{|\det g|}\left(g^{\mu\nu}\partial_\mu\bar\phi\partial_\nu\phi+m^2\bar\phi\phi\right),
\end{equation}
where \(\phi\) is a complex scalar field. When we set the metric to be
\begin{equation}
    g = \alpha\operatorname{diag}(-c^2,1,1,1),
\end{equation}
the \(c\to\infty\) limit either diverges or vanishes or is otherwise trivial regardless of how we scale \(\alpha\) or \(\phi\) or \(m\).

\paragraph{BPS Configuration}
Instead, we now couple the chiral multiplet to a background \(\operatorname U(1)\) gauge supermultiplet \((A_\mu,\chi)\) with charge \(q\) (i.e.\ we consider \(\mathcal N=1\) supersymmetric scalar quantum electrodynamics). The part of the action involving \(\phi\) is now
\begin{equation}
    S = -\int \mathrm d^4x\,\sqrt{|\det g|}\left(g^{\mu\nu}((\partial_\mu+\mathrm iqA_\mu)\bar\phi)((\partial_\nu-\mathrm iqA_\nu)\phi)+m^2\bar\phi\phi\right).
\end{equation}
If we now scale
\begin{align}\label{eq:particle_example_background}
    g_{\mu\nu} &= \operatorname{diag}(-c^2,c^{-2}, c^{-2}, c^{-2}),&
    A_\mu\,\mathrm dx^\mu &= c \mathrm dx^0,&
    (m,q)&\mapsto (c^3 m,c^3 q),&
\end{align}
and split the indices as \((\partial_\mu\phi) = (\dot\phi,\partial_i\phi)\) with \(i\in\{1,2,3\}\) the spatial indices raised and lowered by the Kronecker delta,
we obtain
\begin{equation}
    S = \int \mathrm d^4x\,
    \left(
    c^{-4}\dot{\bar\phi}\dot\phi
    +2\mathrm iq\bar\phi\dot\phi
    - \partial_i\bar\phi \partial_i\phi
    -c^4 (m^2-q^2)\bar\phi\phi
    \right).
\end{equation}
The \(\mathcal O(c^4)\) divergent term cancels if and only if the BPS condition \(m=q\) is satisfied. Assuming this, then in the \(c\to\infty\) limit we obtain the Schrödinger action
\begin{equation}\label{eq:schroedinger-action}
    S = \int \mathrm d^4x\,
    \left(
    2\mathrm iq\bar\phi\dot\phi
    - \partial_i\bar\phi \partial_i\phi
    \right)
\end{equation}
describing a free non-relativistic scalar field \(\phi\), whose excitations can now be interpreted as small fluctuations (of energy \(q+o(1)\)) above a BPS state of energy \(q\) and charge \(q\).

\paragraph{BPS Configuration from Supergravity}
We can regard the above \(\mathcal N=1\) theory with a background supergravity multiplet 
as (a second-quantised\footnote{Since the previous theory was free the particle number is conserved, meaning there is no problem in reducing to a first-quantised description; indeed, after taking the non-relativistic limit \eqref{eq:schroedinger-action}, we obtain an action whose equation of motion, the Schrödinger equation, can be used directly as a first-quantised description of a particle.} description of) a half-BPS line defect coupled to a \(\mathcal N=2\) supergravity with bosonic fields \(g_{\mu\nu},A_\mu\).

In 4d \(\mathcal N=2\) supergravity, a family of BPS black hole solutions are given by the Majumdar--Papapetrou solutions
\begin{align}
    g_{\mu\nu}\,\mathrm dx^\mu\,\mathrm dx^\nu &=
    -H(r)^{-2}\,\mathrm dt^2
    +
    H(r)^2\,\mathrm dx_i\,\mathrm dx^i,&
    A_\mu\,\mathrm dx^\mu &= H^{-1}\,\mathrm dt,
\end{align}
where \(H(x^i)\) is a harmonic function of the transverse coordinates \((x^i)=(x^1,x^2,x^3)\). If we take
\begin{equation}
    H(r) = c^{-1},
\end{equation}
then we obtain the background \((g_{\mu\nu},A_\mu)\) used in \eqref{eq:particle_example_background}. This is an example of the general method outlined above, and it is straightforward to see how one can do this for more general examples.

\section{Construction of Non\texorpdfstring{-}{‐}Relativistic Limits of Maximally Supersymmetric Yang--Mills Theory}\label{sec:construction}
\subsection{Supergravity Backgrounds for Maximally Supersymmetric Yang--Mills Theory}\label{sec:sugra}
We seek non-relativistic limits of four-dimensional maximally supersymmetric Yang--Mills theory, which can be realised as the infrared limit of D3-branes in Type~II string theory. Following the discussion in \cref{sec:review}, therefore, we should couple the D3-brane to appropriate BPS Type~IIB supergravity solutions.
If the supergravity background is simply a D3 black-brane solution with \(\operatorname{Spin}(1,3)\times\operatorname{Spin}(6)\) symmetry, i.e.\ we take a D3NC limit, we obtain plain 4d MSYM \cite{Blair:2023noj} as the infrared limit of Dirac--Born--Infeld theory. In particular, while ten-dimensional Lorentz symmetry is broken, four-dimensional Lorentz symmetry is not broken. In order to take a non-relativistic limit that breaks four-dimensional Lorentz symmetry, we must put on additional charges. An obvious configuration is given by either a D3--F1 or D3--D1 system, consisting of D3-branes along the 0123 directions and either F1 or D1-branes along 04. Applying a general \(\operatorname{SL}(2;\mathbb Z)\) transformation, these are special cases of the D3--\((p,q)\)1 system, for \(p\) and \(q\) coprime integers, where a \((p,q)\)1-brane is a stable bound state of \(p\) F-strings and \(q\) D-strings. In order to not break translation invariance, we smear the \((p,q)\)1-brane along the \(x^i\) and \(Y^A\) directions, as shown in \cref{table:brane-arrangement}.
\begin{table}
\begin{center}
\begin{tabular}{rcccccccccc}\toprule
&0&1&2&3&4&5&6&7&8&9\\
&\(t\)&\(x^1\)&\(x^2\)&\(x^3\)&\(X\)&\(Y^1\)&\(Y^2\)&\(Y^3\)&\(Y^4\)&\(Y^5\)\\
\midrule
D3&\(\bullet\)&\(\bullet\)&\(\bullet\)&\(\bullet\)\\
\((p,q)\)1&\(\bullet\)&---&---&---&\(\bullet\)&---&---&---&---&---\\
\bottomrule
\end{tabular}
\end{center}
\caption{Quarter-BPS brane configuration in Type~IIB string theory for taking non-relativistic limits of four-dimensional maximally supersymmetric Yang--Mills theory. The symbol `\(\bullet\)' denotes that a brane extends along a given direction, and the symbol `---' denotes that a brane is smeared along a given direction.}\label{table:brane-arrangement}
\end{table}
Adapted to this brane configuation, we parameterise \(\mathbb R^{1,9}\) in terms of these \((t,x^i,X,Y^A)\) where \(i\in\{1,2,3\}\) and \(A\in\{1,\dotsc,5\}\). These indices are raised and lowered with the Kronecker delta. It was shown in \cite{Lambert:2024ncn} that the joint limit associated with a quarter-BPS brane configuration can be implemented sequentially with no subtleties; we shall therefore consider the decoupling limit associated with a \((p,q)\)1-brane and apply it directly to 4d MSYM.

We start with the following solution to Type~IIB supergravity describing a (smeared) F1-string along \((t,X)\):
\begin{equation}\label{eq:sugra-solution}
\begin{aligned}
    G &= H^{-1}\,(-\mathrm dt^2+\mathrm dX^2)+\mathrm dx_i\,\mathrm dx^i+\mathrm dY_A\,\mathrm dY^A, \\
    B_2 &= H^{-1} \,\mathrm dt\wedge\mathrm dX,\\
    \exp\Phi &= g_\mathrm s H^{-1/2} ,
\end{aligned}
\end{equation}
where \(H\) is a harmonic function of \(x^i\) and \(Y^A\). Note that we are not including a flat contribution to $B_2$ here. We specialise \eqref{eq:sugra-solution} to the case where \(H=c^{-2}\) is constant to obtain
\begin{subequations}\label{eq:sugra-constant-solution}
\begin{align}
    G &= c^2 (-\mathrm dt^2 + \mathrm dX^2) + \mathrm dx_i \mathrm dx^i + \mathrm dY_A\,\mathrm dY^A, \\
    B_2 &= c^2 \mathrm dt \wedge\mathrm dX,\\
    \exp\Phi &= g_\mathrm s c .
\end{align}
\end{subequations}
Here, \(c\) is an arbitrary dimensionless constant parameter that we will eventually send to infinity (but is finite for now). Since we are only interested in its limit we can absorb the string coupling through a redefinition of $c$ combined with a rescaling of $t$ and $X$. We shall assume this from here onwards, effectively setting $g_\mathrm s = 1$ throughout.

The configuration \eqref{eq:sugra-constant-solution} is a nonsingular (indeed, flat) solution to the supergravity equations of motion, so any \(\operatorname{SL}(2;\mathbb R)\) transformation parameterised by
\begin{equation}
    \begin{pmatrix}
        \alpha&  \beta\\  \gamma&\delta
    \end{pmatrix}\in\operatorname{SL}(2;\mathbb R)
\end{equation}
will generate another solution to Type~IIB supergravity \cite{Schwarz:1995dk}, namely
\begin{subequations}\label{eq:transformed-sugra-solution}
\begin{align}
G&= \sqrt{\gamma^2 + c^2 \delta^2} \left(c(-\mathrm dt^2+\mathrm dX^2) + c^{-1}\left(\mathrm dx_i\,\mathrm dx^i+\mathrm dY_A\,\mathrm dY^A\right)\right) \ ,\\
\exp\Phi&=c^{-1}(\gamma^2+c^2\delta^2) \ ,\\
B_2&= \delta c^2 \,\mathrm dt\wedge\mathrm dX \ ,\\
C_0&= \frac{\alpha\gamma+c^2\beta\delta}{\gamma^2+c^2\delta^2} \ ,\\
C_2&= -\beta c^2 \,\mathrm dt\wedge\mathrm dX \ .
\end{align}
\end{subequations}
Under the \(\operatorname{SL}(2;\mathbb Z)\)-valued discrete U-duality of string theory, this will only be physically equivalent to the original one if \(\alpha,\beta,\gamma,\delta\in\mathbb Z\).
We assume that the parameters \((\alpha,\beta,\gamma,\delta)\) do not scale with respect to \(c\).

\paragraph{NS--NS Couplings}
Ignoring the couplings to the Ramond--Ramond fields \(C_2\) and \(C_0\), the D3-brane worldvolume action couples to the NS--NS fields as
\begin{align} \nonumber
    \hat S_\mathrm{YM} = \int\frac{\mathrm dt\,\mathrm d^3x}{4\pi} \tr\bigg(& 
    \frac{F_{0i} F_{0i}}{\gamma^2 + \delta^2 c^2} + \frac{2 c^2 \delta}{\gamma^2 + \delta^2 c^2} D_i X F_{0i} - \frac{c^2}{2(\gamma^2 + \delta^2 c^2)} F_{ij} F_{ij} \\ \nonumber
    &+ (D_0 X)^2 + \Big( \frac{\delta^2 c^4}{\gamma^2 + \delta^2 c^2} - c^2 \Big) D^i X D_i X \\  \nonumber
    & + 2\mathrm i\delta [X, Y^A] D_0 Y^A - D_i Y^A D_i Y^A + \gamma^2 [X,Y^A][X,Y^A] \\ \label{eq:D3-action-before-limit} 
    & + \delta^2 [Y^A, Y^B][Y^A , Y^B] +O(c^{-2}) \bigg) \ ,
\end{align}
where we have used the relation
\begin{equation}
    (2\pi\alpha')^2 T_3 = \frac{1}{2\pi}
\end{equation}
between the D3-brane tension \(T_3\) and the Regge slope \(\alpha'=\ell_{\mathrm s}^2\) (recall that we used the scaling of $c$ to set $g_\mathrm s = 1$).

\paragraph{Wess--Zumino Terms} 
The Ramond--Ramond axion fields \(C_0\) and \(C_2\) couple via the topological Wess--Zumino terms
\begin{equation} \label{eq: general WZ terms}
    \hat S_\mathrm{WZ} = \frac{1}{2\pi} \int \bigg( \frac12 C_0\tr(F\wedge F) + 
    C_0\tr(\Tilde{B}_2 \wedge F)+\tr(\Tilde{C}_2\wedge F) \bigg) \ .
\end{equation}
We use the notation $\Tilde{B}_2, \Tilde{C}_2$ to denote the relevant non-Abelian pull-back of the two-form fields to the D3-brane worldvolume; as each contains a single appearance of $X$, this includes a factor of $2\pi\alpha'$ that we have pulled out to match normalisations from the spacetime coordinate to worldvolume scalar field.
Since \(B_2\) and \(C_2\) given in \eqref{eq:transformed-sugra-solution} both pull back to forms proportional to $D_i X\,\mathrm dt \wedge\mathrm dx^i$, we have \(\Tilde{B}_2\wedge \Tilde{B}_2 = \Tilde{B}_2\wedge \Tilde{C}_2=0 \) and have therefore ignored terms including these in \eqref{eq: general WZ terms}.

The \(C_0\) term is
\begin{align} \nonumber
    \hat S_{\mathrm{WZ},C_0} &= \frac{1}{4\pi} \int C_0 \tr(F \wedge F)
    \\ \label{eq:wess-zumino-term}
    &= \frac{\alpha\gamma+c^2\beta\delta}{4\pi(\gamma^2+c^2\delta^2)}\int\tr(F\wedge F),
\end{align}
which we recognise as the $\theta$-term
\begin{align}
    S_{\theta} =  \frac{\theta}{8\pi^2} \int \tr( F\wedge F) \ ,
\end{align}
with 
\begin{align}\label{eq:theta-angle}
    \frac{\theta}{2\pi}=  \frac{\alpha\gamma+c^2\beta\delta}{\gamma^2+c^2\delta^2} .
\end{align}
As for the other terms, we see that they combine to make
\begin{equation} \label{eq: 2-form WZ terms}
    \hat S_{\mathrm{WZ}, 2} =  -c^2 \int\frac{\mathrm dt\,\mathrm d^3 x}{2\pi}    \, \left(\beta 
    - \frac{\delta\left(\alpha \gamma + c^2 \beta \delta\right)}{\gamma^2 + \delta^2 c^2} \right) \varepsilon^{ijk} \tr(D_i X F_{jk} ) \ .
\end{equation}
Since the denominators in \eqref{eq: 2-form WZ terms} are \(\gamma^2+\delta^2c^2\), same as the denominators in \eqref{eq:D3-action-before-limit} and \eqref{eq:wess-zumino-term}, the qualitative behaviour of the \(c\to\infty\) limit depends on whether \(\delta\ne0\). We discuss these two cases separately in \cref{sec:d1nc} (for \(\delta=0\)) and \cref{sec:snc} (for \(\delta\ne0\)). However, it is worth mentioning briefly that the leading terms in \eqref{eq: 2-form WZ terms} cancel when $\delta\neq0$, leaving us with a finite Wess-Zumino term.

\subsection{The Generalised D1NC Limit} \label{sec:d1nc}

We first consider the case where \(\delta=0\). This case will turn out to be the same as the ordinary D1NC limit of 4d MSYM (as discussed in \cite{Lambert:2024yjk}) together with a topological \(\theta\)-term, so we call it the `generalised D1NC limit'.

In the case \(\delta=0\), we have \(\beta=-1/\gamma\) with \(\alpha\) arbitrary, and the supergravity background is (taking $\gamma>0$)
\begin{subequations} \label{eq: generalised D1NC limit 1}
\begin{align}
    G_{10}&= \gamma(c(-\mathrm dt^2+\mathrm dX^2) + c^{-1}\left((\mathrm dx^i)^2+(\mathrm dY^A)^2\right) \ ,\\
    \exp\Phi&=c^{-1}\gamma^2 \ ,\\
    B_2&=0 \ ,\\
    C_0&= \alpha/\gamma \ ,\\
    C_2&= \gamma^{-1} c^2\,\mathrm dt\wedge\mathrm dX \ .
\end{align}
\end{subequations}
After a rescaling of $c$ to normalise the dilaton to $\exp\Phi = c^{-1}$, followed by an additional rescaling of our coordinates, this becomes
\begin{subequations} \label{eq: generalised D1NC limit 2}
\begin{align}
    G_{10} &=  c \left(-\mathrm dt^2 +\mathrm dX^2\right) + c^{-1}\left(\mathrm dx_i\mathrm dx^i+\mathrm dY_A\,\mathrm dY^A\right) \ ,\\
    \exp\Phi &= c^{-1} \ ,\\
    B_2 &= 0 \ ,\\
    C_0 &= \alpha / \gamma \ ,\\
    C_2 &= c^2\,\mathrm dt \wedge\mathrm dX \ ,
\end{align}
\end{subequations}
which is the same as the usual D1NC background \cite{Lambert:2024yjk} but with an extra \(C_0\) turned on. If we restrict to $\operatorname{SL}(2;\mathbb{Z})$ transformations then we have $\gamma = - \beta = 1$ with $\alpha \in \mathbb{Z}$ a free parameter, meaning the two solutions \eqref{eq: generalised D1NC limit 1} and \eqref{eq: generalised D1NC limit 2} are identical aside from a rescaling of $\alpha$.

Indeed, with \(\delta=0\), the action \eqref{eq:D3-action-before-limit} simplifies to
\begin{align} \nonumber
    \hat{S}_{YM} =   \int\frac{\mathrm dt\,\mathrm d^3 x}{4\pi} \, \tr\bigg(&
    \frac{F_{0i} F_{0i}}{\gamma^2} - \frac{c^2}{2\gamma^2} F_{ij} F_{ij}
    + (D_0 X)^2 - c^2 D_i X D_i X \\
    &- D_i Y^A D_i Y^A
    + \gamma^2 [X,Y^A]^2 + O(c^{-2}) \bigg) \ ,
\end{align}
augmented by the divergent two--form Wess--Zumino term
\begin{equation}
    \hat{S}_{WZ,2} = - c^2 \int \frac{ \mathrm dt \mathrm d^3 x}{4\pi} \, \beta \varepsilon^{ijk} \tr ( D_i X F_{jk} ) \ ,
\end{equation}
as well as the finite $\theta$-term
\begin{equation}
    \hat{S}_{\theta} = \frac{\alpha}{\gamma} \int \frac{\mathrm dt\, \mathrm d^3 x}{4\pi} \,\varepsilon_{ijk} \tr ( F_{0i} F_{jk} ) \ ;
\end{equation}
following the discussion of \cite{Lambert:2024yjk}, we can use a Hubbard--Stratonovich transformation to regulate the divergent terms and take the $c\to\infty$ limit, obtaining the bosonic action
\begin{align} \nonumber
    S_{D1} = \int \frac{ \mathrm dt \mathrm d^3 x}{4\pi} \tr\Big(& \gamma^{-2} F_{0i} F_{0i} + D_0 X D_0 X - G_{ij} \left( F_{ij} - \gamma \varepsilon_{ijk} D_k X \right) \\ \label{eq: d1nc pre normalisation}
    &- D_i Y^A D_i Y^A + \gamma^2 [X,Y^A]^2 + \alpha \gamma^{-1} \varepsilon_{ijk} F_{0i} F_{jk} \Big) \ .
\end{align}
One may rescale the fields of the theory by
\begin{align}
    X &\mapsto \gamma^{-1} X \ , &
    Y^A &\mapsto \gamma^{-1} Y^A \ , &
    G_{ij} &\mapsto \gamma^{-2} G_{ij} \ ,
\end{align}
to rewrite the action as
\begin{align} \nonumber
    S_{D1} = \, &\int \frac{ \mathrm dt \mathrm d^3 x}{ \gd^2} \Big( F_{0i} F_{0i} + D_0 X D_0 X - G_{ij} \left( F_{ij} -  \varepsilon_{ijk} D_k X \right) \\ \nonumber
    &\hspace{2.0cm}- D_i Y^A D_i Y^A + [X,Y^A]^2 \Big) \\ \label{eq: gD1 action}
    &+ \frac{\theta_{\mathrm{D1NC}}}{8 \pi^2} \int\mathrm dt\,\mathrm d^3 x\, \varepsilon_{ijk} \tr (F_{0i} F_{jk} )  \ ,
\end{align}
where we have defined the coupling
\begin{equation}
    \gd^2 = 4\pi \gamma^2 \ ,
\end{equation}
and, as above, the $\theta$ angle
\begin{equation}
    \frac{\theta_{\mathrm{D1NC}}}{2\pi} = \frac{\alpha}{\gamma} \ .
\end{equation}
The interesting dynamics of the theory occur once we give $X$ a VEV and allow for non-trivial spatial profile, and so we must be somewhat careful to take into account its rescaling by $\gamma$ when comparing theories before and after $\operatorname{SL}(2;\mathbb{Z})$ transformations. Similarly, while we chosen to write the action in the form \eqref{eq: gD1 action} we could remove the overall factor of $\gd^2$ using the rescaling
\begin{equation}
\begin{aligned}
    t &\mapsto \gd^{-1} t \ , &
    x^i &\mapsto \gd x^i \ , \\
    X &\mapsto \gd^{-1} X \ , &
    Y^A &\mapsto \gd Y^A \ , &
    G_{ij} &\mapsto \gd^2 G_{ij} \ ,
\end{aligned}
\end{equation}
at the expense of rescaling the VEV of $X$. In order to fix this ambiguity, we choose to fix $\langle X \rangle$ with respect to the normalisation in \eqref{eq: gD1 action}, in which case $\gd$ becomes meaningful.

The couplings \(\theta_\mathrm{D1NC}\) and \(g_\mathrm{D1NC}\) determine a unique element of \(\operatorname{PSL}(2;\mathbb R)\); fixing the signs of parameters by taking $\gamma>0$, this is given by
\begin{equation}\label{eq:d1nc-parameterisation}
    \begin{pmatrix}\alpha& \beta\\\gamma&\delta\end{pmatrix}
    = \frac{\gd}{\sqrt{4\pi}}
    \begin{pmatrix}
        \theta_{\mathrm{D1NC}}/2\pi & - 4\pi/\gd^2\\
         1 & 0
    \end{pmatrix} \ ,
\end{equation}
where the coupling $\gd$ is by definition positive.

\subsection{The Generalised SNC Limit}\label{sec:snc}
The more general case to consider is \(\delta\ne0\). This case will turn out to be the same as the ordinary SNC limit of 4d MSYM (which was discussed in \cite{Fontanella:2024rvn}), i.e.\ the supersymmetric version of the Galilean Yang--Mills theory \cite{Bagchi:2015qcw}, but with two different kinds of topological terms, so we call this case the `generalised SNC limit'.
(Recall that the four-dimensional Galilean Yang--Mills theory is the null reduction of five-dimensional Yang--Mills theory \cite{Bagchi:2022twx}.)

When \(\delta\ne0\), there are no terms in \eqref{eq:D3-action-before-limit} that scale with positive powers of $c$, and so we find
\begin{align} \nonumber
    S_\mathrm{SNC,YM} = \int \frac{\mathrm dt\,\mathrm d^3x}{{4\pi}}\,\tr \bigg(& 
    (D_0 X)^2 + \frac{2}{\delta} D_i X F_{0i} - \frac1{2 \delta^2} F_{ij} F_{ij} - \frac{\gamma^2}{\delta^2} D_i X D_i X  \\ \nonumber
    &+ 2\mathrm i\delta [X,Y^A] D_0 Y^A - D_i Y^A D_i Y^A  \\
    &+  \gamma^2 [X, Y^A][X, Y^A] + \frac{\delta^2}{2} [Y^A , Y^B][Y^A , Y^B] \bigg) \ ,
\end{align}
in the \(c\to\infty\) limit. As discussed in section \ref{sec:sugra}, when taking the \(c\to\infty\) limit of the two-form Wess-Zumino terms \eqref{eq: 2-form WZ terms} the divergent pieces cancel internally, leaving the finite terms
\begin{equation}
    S_\mathrm{SNC,WZ} = \int \frac{\mathrm dt \, \mathrm d^3x}{4 \pi} \, \varepsilon^{ijk} \tr\left(\frac\beta{\delta} F_{0i}F_{jk}
    + \frac\gamma{\delta^2} D_iXF_{jk} \right),
\end{equation}
where we have used \(\alpha\delta - \beta \gamma = 1\) to reach this final form. Together these form
\begin{align} \nonumber
    S_{\mathrm{SNC}} = \int \frac{\mathrm dt\, \mathrm d^3 x}{4\pi}\tr \bigg(& (D_0 X)^2 + \frac{2}{\delta} D_i X F_{0i} - \frac{1}{2\delta^2} ( F_{ij} - \gamma \varepsilon_{ijk} D_k X)^2 \\ \nonumber
    &+ 2 i \delta [X,Y^A] D_0 Y^A - (D_i Y^A)^2 + \gamma^2 [X,Y^A]^2 \\ \label{eq: gsnc action pre-rescaling}
    &+ \frac{\delta^2}{2} [Y^A, Y^B]^2 + \frac{\beta}{\delta} \varepsilon^{ijk} F_{0i} F_{jk} \bigg) \ .
\end{align}
We can clean this up using the rescalings
\begin{align}
    X &\mapsto \delta^{-1} X \ , &
    Y^A &\mapsto \delta^{-1} Y^A
\end{align}
to reach
\begin{align} \nonumber
    S_{\mathrm{SNC}} =\,& \int \frac{\mathrm dt\, \mathrm d^3 x}{\gsnc^2}\tr \bigg( (D_0 X)^2 + 2 D_i X F_{0i} - \frac{1}{2} ( F_{ij} - \lambda \varepsilon_{ijk} D_k X)^2 \\ \nonumber
    &\hspace{2.2cm}+ 2 i  [X,Y^A] D_0 Y^A - (D_i Y^A)^2 \\ \nonumber
    &\hspace{2.2cm}+ \lambda^2 [X,Y^A]^2 + \frac{1}{2} [Y^A, Y^B]^2 \bigg) \\ \label{eq: gen SNC action}
    &+ \frac{\theta_{\mathrm{SNC}}}{8\pi^2} \int \mathrm dt\,\mathrm d^3 x \, \varepsilon_{ijk} \tr( F_{0i} F_{jk}) \ ,
\end{align}
where we have defined
\begin{subequations} \label{eq: snc parameter def}
\begin{align}
    \gsnc^2 &= 4\pi \delta^2 \ , \\
    \lambda &= \gamma \delta^{-1} \ , \\
    \frac{\theta_{\mathrm{SNC}}}{2\pi} &= \beta \delta^{-1} \ .
\end{align}
\end{subequations}
We can remove the coupling $\gsnc$ from the action via the rescalings
\begin{align}
    t &\mapsto \gsnc^2 t \ , &
    X &\mapsto \gsnc^2 X \ ,
\end{align}
at the expense of a rescaling of $\lambda$ and the VEV of $X$. Like in the D1NC limit the VEV is responsible for the dynamics of the theory \cite{Lambert:2026}; we again choose to fix its value using the convention \eqref{eq: gen SNC action}, in which case the coupling $\gsnc$ becomes a meaningful quantity.

As we observed in the previous section for the generalised D1NC limit, the couplings in the action define a unique element of $\operatorname{PSL}(2;\mathbb{R})$; fixing the overall sign by taking $\delta > 0$, this is
\begin{equation}\label{eq:snc-parameterisation}
    \begin{pmatrix}\alpha& \beta\\\gamma&\delta\end{pmatrix}
    = \frac{\gsnc}{\sqrt{4\pi}}
    \begin{pmatrix}
         4\pi/ \gsnc^2 + \lambda \theta_{\mathrm{SNC}} / 2\pi &  \theta_{\mathrm{SNC}} /2\pi \\
         \lambda & 1
    \end{pmatrix} \ ,
\end{equation}
where similarly to the above we are taking $\gsnc>0$.

While this looks somewhat different to the standard  supersymmetric Galilean Yang-Mills action, it is straightforward to see that the two are equivalent to up to boundary terms using the redefinition
\begin{equation}
    A_0 \mapsto A_0 - \frac{\lambda^2}{2} X
\end{equation}
of $A_0$, for which the action becomes
\begin{align} \nonumber
    S_{\mathrm{SNC}} =\,& \int \frac{\mathrm dt\, \mathrm d^3 x}{\gsnc^2} \tr \bigg( 
    (D_0 X)^2 + 2 D_i X F_{0i} - \frac{1}{2} F_{ij} F_{ij} + 2 i [X, Y^A] D_0 Y^A \\ \nonumber
    &\hspace{2.2cm}- (D_i Y^A)^2 + \frac{1}{2} [Y^A , Y^B]^2 + \Lambda \varepsilon_{ijk} D_i X F_{jk} \bigg) \\
    &+ \frac{\theta_{\mathrm{SNC}}}{8\pi^2} \int\mathrm dt\,\mathrm d^3 x \, \varepsilon_{ijk} \tr( F_{0i} F_{jk}) \ .
\end{align}
Here we have defined
\begin{equation}
    \Lambda = \lambda + \frac{\gsnc^2 \lambda^2 \theta_{\mathrm{SNC}}}{16\pi^2} =\frac{\alpha^2 \delta^2 - 1}{2 \beta \delta} \ .
\end{equation}
In other words, the deformation is equivalent to including a $\theta$-term and a term proportional to the monopole charge. As an aside, it is straightforward to see that this arises from a flat SNC limit with background Ramond--Ramond fields:
\begin{subequations}
\begin{align}
    G_{10} &= c^2 (-\mathrm dt^2 + \mathrm dX^2) + \mathrm dx^2 + \mathrm dY^2 \ , \\
    \exp\Phi &= \delta^2 c \ , \\
    B_2 &= c^2\,\mathrm dt \wedge\mathrm dX  \ , \\
    C_0 &= \frac{\beta}{\delta} \ , \\
    C_2 &= - \frac{\beta c^2}{\delta}\mathrm dt \wedge\mathrm dX + \frac{\alpha^2 \delta^2 - 1}{2 \beta \delta^3} \,\mathrm dt \wedge\mathrm dX \ .
\end{align}
\end{subequations}
Since the divergent part of $C_2$ is $-C_0 B_2^{(\text{div.})}$, this is indeed an SNC limit, as pointed out in \cite{Bergshoeff:2022iss, Ebert:2023hba}.

\section{S-duality For Non-Relativistic Limits}\label{sec:duality}

\subsection{Dualities From the Abelian Path Integral}

We have seen how to construct two classes of theory by taking non-Lorentzian limits of $\mathcal{N}=4$ SYM related by $\operatorname{SL}(2;\mathbb{R})$ transformations. It is well-established that the $\operatorname{SL}(2;\mathbb{Z})$ subgroup of these transformations acts as a duality on the relativistic theory; it therefore seems natural to suppose that this extends to the non-relativistic limits, with the same subgroup mapping us between theories either within each class or between the two. Such a proposal is obviously difficult to study in the non-Abelian theory, and we shall first look at the Abelian limit to gain some intuition for the problem.

Recall that electric--magnetic duality of Maxwell theory can be shown directly via a path-integral computation in which a Lagrange multiplier implementing the Bianchi identity becomes the dual potential \cite{Deser:1976iy}, though this derivation misses certain nonperturbative considerations \cite{Witten:1995gf,Donnelly:2016mlc}, cf.\ \cite{Borsten:2025phf}. We similarly directly show that a duality between the generalised D1NC and SNC theories in the case of an Abelian gauge group. Of course, in these cases the theories are trivial unless coupled to new charged matter fields, but we shall neglect this in what follows. We start with the SNC theory. Since the $Y^A$ scalar fields completely decouple from the gauge sector in the Abelian theory (and are already equivalent to the corresponding fields in the Abelian D1NC theory) we will ignore them. As with standard electric--magnetic duality, we can work purely in terms of field strengths by introducing a Lagrange multiplier field $B_{\mu}$ into the action,
\begin{equation}
    S_{GEM} = \frac{1}{2g^2} \int \mathrm dt\,\mathrm d^3 x \bigg( (\partial_0 X)^2 + 2 \partial_i X F_{0i} - \frac{1}{2} F_{ij} F_{ij} + \frac{g^2}{4\pi} \varepsilon^{\mu\nu\rho\sigma} \partial_{\mu} B_{\nu} F_{\rho\sigma} \bigg) \ ,
\end{equation}
where the equation of motion for $B$ imposes the Bianchi identity\footnote{As usual, we introduce an additional factor of $2$ between the Abelian and non-Abelian couplings to account for the normalisation of the trace in the non-Abelian action.}. Expanding this out gives
\begin{align} \nonumber
    S_{GEM} = \frac{1}{2g^2} \int\mathrm dt\,\mathrm d^3 x \bigg(& (\partial_0 X)^2 + 2 \Big( \partial_i X - \frac{g^2}{4\pi} \varepsilon_{ijk} \partial_j B_k \Big) F_{0i} \\
    &-\frac{1}{2} F_{ij} F_{ij} - \frac{g^2}{4\pi} \varepsilon_{ijk} F_{ij} \Big( \partial_0 B_k - \partial_k B_0 \Big) \bigg) \ .
\end{align}
Defining
\begin{subequations}
\begin{align}
    \Tilde{F}_{0i} &= \partial_0 B_i - \partial_i B_0 \ , \\
    \Tilde{F}_{ij} &= 2 \partial_{[i} B_{j]} \ , \\ \label{eq: X rescaling}
    \Tilde{X} &= 4\pi g^{-2} X \ ,
\end{align}
\end{subequations}
this can be rewritten as
\begin{align} \nonumber
    S_{GEM} = \frac{g^2}{2(4\pi)^2} \int\mathrm dt\,\mathrm d^3 x \bigg(& (\partial_0 \Tilde{X})^2 + \frac{4\pi}{g^{2}} \varepsilon_{ijl} F_{0l} \Big( \varepsilon_{ijk} \partial_k \Tilde{X} - \Tilde{F}_{ij} \Big)  \\
    &-\frac{(4\pi)^2}{2 g^4} \Big( F_{ij} + \frac{g^2}{4\pi} \varepsilon_{ijk} \Tilde{F}_{0k} \Big)^2  + \Tilde{F}_{0i} \Tilde{F}_{0i} \bigg) \ .
\end{align}
As we are using a Lagrange multiplier to enforce the Bianchi identity, the path integral is over $F_{ij}$ and $F_{0i}$ as independent fields. $F_{ij}$ appears quadratically, and so we can integrate it out at the expense of a overall constant that we drop as in the derivation of electromagnetic duality for Maxwell theory. However, $F_{0i}$ appears linearly: it therefore acts as a Lagrange multiplier that imposes the constraint 
\begin{equation}
    \Tilde{F}_{ij} = \varepsilon_{ijk} \partial_k \Tilde{X} \ .
\end{equation}
Further defining $\Tilde{g}^2 = (4\pi)^2 g^{-2}$ and $G_{ij} = \tilde{g}^2 (4\pi)^{-1} \varepsilon_{ijk} F_{0k}$, we obtain the path-integral identity
\begin{align}
    \int \mathrm D A\,\mathrm D X \,\exp(\mathrm iS_{GEM}) = \int\mathrm DB\,\mathrm D\Tilde{X}\,\mathrm DG  \,\exp(\mathrm i S_{D1})\ ,
\end{align}
where
\begin{equation}
    S_{D1} = \frac{1}{2 \Tilde{g}^2} \int\mathrm dt\,\mathrm d^3 x \bigg( (\partial_0 \Tilde{X})^2 + \Tilde{F}_{0i} \Tilde{F}_{0i} - G_{ij} \Big( \Tilde{F}_{ij} - \varepsilon_{ijk} \partial_k \Tilde{X} \Big) \bigg)
\end{equation}
is the Abelian D1NC action with gauge field $B_{\mu}$. We can also work in reverse: starting from the D1NC theory and introducing $B_{\mu}$ we see that $F_{ij}$ acts as a Lagrange multiplier imposing $\varepsilon_{ijk} \Tilde{F}_{0k} = G_{ij}$, and that the path integral over $F_{0i}$ gives us a $\Tilde{F}_{ij}^2$ term. Putting this together, we recover the SNC action.

Let us now see how this changes when we go to the generalised theories, which for reference are
\begin{align} \nonumber
    S_{\mathrm{gGEM}} = \int \frac{\mathrm dt\,\mathrm d^3 x}{2g^2} \bigg(&
    (\partial_0 X)^2 + 2 \partial_i X F_{0i} - \frac{1}{2} (F_{ij} - \lambda \varepsilon_{ijk} \partial_k X)^2 + \frac{g^2 \theta}{8 \pi^2} \varepsilon_{ijk} F_{0i} F_{jk} \\
    &- \frac{g^2}{4\pi} \varepsilon_{ijk} F_{0i} \Tilde{F}_{jk} - \frac{g^2}{4\pi} \varepsilon_{ijk} \Tilde{F}_{0i} F_{jk} \bigg) \ .
\end{align}
and
\begin{align} \nonumber
    S_{\mathrm{gD1}} = \int \frac{\mathrm dt\,\mathrm d^3 x}{2 g^2} \bigg(& (\partial_0 X)^2 + F_{0i}^2 - G_{ij} ( F_{ij} - \varepsilon_{ijk} \partial_k X) + \frac{g^2 \theta}{8\pi^2} \varepsilon_{ijk} F_{0i} F_{jk} \\
    & + \frac{g^2}{4\pi} \varepsilon_{ijk} F_{0i} \Tilde{F}_{jk} + \frac{g^2}{4\pi} \varepsilon_{ijk} \Tilde{F}_{0i} F_{jk} \bigg) \ ,
\end{align}
where we have chosen opposite sign conventions for the Lagrange multiplier terms in each for later convenience; as the theories are free, this will not affect our discussion. The analysis for the generalised D1NC theory is fairly straightforward: $F_{ij}$ appears linearly and imposes the constraint
\begin{equation}
    G_{ij} = \frac{g^2}{4\pi} \varepsilon_{ijk} \Tilde{F}_{0k} + \frac{g^2\theta}{8\pi^2} \varepsilon_{ijk} F_{0k} \ ,
\end{equation}
which fixes the value of $G_{ij}$. Performing the integral over $F_{0i}$ and rescaling $X$ as in \eqref{eq: X rescaling} leaves us with the generalised SNC theory with parameters
\begin{align}
    \Tilde{g} &= \frac{4\pi}{g} \ , &
    \Tilde{\theta} &= 0 \ , &
    \Tilde{\lambda} &= -\frac{g^2 \theta}{8\pi^2} \ .
\end{align}
Things are more interesting in the generalised SNC theory. Just as in the standard theory $F_{0i}$ appears linearly here, meaning we may integrate it out to impose the constraint
\begin{equation} \label{eq: abelian duality condition}
     \Tilde{F}_{ij} =  \frac{\theta}{2\pi} F_{ij} + \frac{4\pi}{g^2} \varepsilon_{ijk} \partial_k X  \ .
\end{equation}
The resulting theory then depends on whether $\theta$ vanishes. If it does, things proceed largely as above: we impose the constraint with a Lagrange multiplier, rescale $X$, and redefine $G_{ij}$ by an additive term to obtain the generalised D1NC theory with the couplings
\begin{align}
    \Tilde{g} &= \frac{4\pi}{g} \ , &
    \Tilde{\theta} &= - \frac{ g^2 \lambda}{2} \ .
\end{align}
However, when $\theta\neq0$ the constraint \eqref{eq: abelian duality condition} fixes the value of $F_{ij}$, meaning there is no additional Lagrange multiplier that needs to be introduced. We therefore remain with the generalised SNC limit, and after rescaling $X$ to
\begin{equation}
    X = \frac{2\pi}{\theta} \Tilde{X}
\end{equation}
we recover $S_{\mathrm{gGEM}}$ with the parameters
\begin{align}
    \Tilde{g}^2 &= \frac{g^2\theta^2}{4\pi^2} \ , &
    \Tilde{\theta} &= -\frac{4\pi^2}{\theta} \ , &
    \Tilde{\lambda} &=   \lambda + \frac{8\pi^2}{g^2 \theta}  \ .
\end{align}
This structure, in which the theory we find depends on the value of $\theta$, was also observed in \cite{Blair:2025prd}, in which the axiodilaton-dependence of S-duality within non-Lorentzian 1-brane limits of type IIB supergravity was studied.

\subsection{The Action of \texorpdfstring{$\operatorname{SL}(2;\mathbb Z)$}{SL(2; ℤ)}}

The two theories constructed in this work, namely the generalised D1NC theory (parameterised by \((g_\mathrm{D1NC},\theta_\mathrm{D1NC})\)) and the generalised SNC theory (parameterised by \((g_\mathrm{SNC},\theta_\mathrm{SNC},\lambda) \)), possess moduli spaces that fit together into a space diffeomorphic to \(\operatorname{PSL}(2;\mathbb R)\), as given in \eqref{eq:d1nc-parameterisation} and \eqref{eq:snc-parameterisation}. We have seen that the Abelian theories map into each other under an S-duality transformation, with the transformation rules
\begin{subequations} \label{eq: S dualities}
\begin{align} \label{eq: d1nc s duality}
    \operatorname{D1NC}(g,\theta)
    &\overset{\mathsf S } \mapsto \operatorname{SNC}(4\pi/g,0,-g^2\theta/8\pi^2),\\
    \operatorname{SNC}(g,\theta,\lambda)&\overset{\mathsf S}\mapsto\begin{cases}
    \operatorname{D1NC}(4\pi/g,-g^2 \lambda/2)&\text{if \(\theta=0\)},\\ \label{eq: SNC S duality}
    \operatorname{SNC}\mleft(g|\theta|/2\pi,-4\pi^2/\theta,\lambda + 8\pi^2 / g^2 \theta \mright)&\text{if \(\theta\ne0\)}, \end{cases}
\end{align}
\end{subequations}
(we must also take into account the transformations of the VEV, though will not notate that explicitly here). Let us assume that this structure remains in the non-Abelian theories, with evidence for this claim presented in the next section. As with the relativistic theory, upon Wick-rotating the $\theta$-terms, we see that on a $k$-instanton configuration it evaluates to
\begin{equation}
    \exp(- S^{(E)}_{\theta}) = \exp(\mathrm i k \theta) \ ,
\end{equation}
meaning the theory's dynamics can only sees the periodic variable $\theta \sim \theta + 2 \pi$; we therefore also have duality symmetries that shift the angle by its period,
\begin{subequations} \label{eq: t dualities}
\begin{align}
    \operatorname{D1NC}(g,\theta)&\overset{\mathsf T}\mapsto \operatorname{D1NC}(g,\theta+2\pi) \ ,\\
    \operatorname{SNC}(g,\theta,\lambda)&\overset{\mathsf T}\mapsto \operatorname{SNC}(g,\theta+2\pi,\lambda) \ .
\end{align}
\end{subequations}
As is well--known, together these generate the group $\operatorname{PSL}(2;\mathbb{Z})$. 

The most natural action of this group on the moduli space is via left or right matrix multiplication. It is straightforward to see that the left action generates the dualities \eqref{eq: S dualities} and \eqref{eq: t dualities}. Concretely, let us define the generators of the left action of \(\operatorname{PSL}(2;\mathbb Z)\) as
\begin{subequations}
\begin{align}
    \mathsf S &= \begin{pmatrix}
        0 & -1 \\
        1 & 0        
    \end{pmatrix}, \\
    \mathsf T &= \begin{pmatrix}
        1 & 1 \\
        0 & 1
    \end{pmatrix}.
\end{align}
\end{subequations}
The action of an $\mathsf S$ transformation on a generalised SNC theory's matrix representation \eqref{eq:snc-parameterisation} is then
\begin{align} \nonumber
    \mathsf S \circ \mathrm{M}_{\mathrm{gSNC}} &= \frac{g}{\sqrt{4\pi}}
    \begin{pmatrix}
        0 & -1 \\
        1 & 0
    \end{pmatrix} 
    \begin{pmatrix}
         4\pi/ g^2 + \lambda \theta / 2\pi &  \theta /2\pi \\
         \lambda & 1
    \end{pmatrix} \\
    &= \frac{g}{\sqrt{4\pi}} \begin{pmatrix}
        - \lambda & -1 \\
         4\pi / g^2 + \lambda \theta / 2 \pi & \theta/2\pi
    \end{pmatrix} \ .
\end{align}
If $\theta=0$, we recognise this as the generalised D1NC moduli matrix \eqref{eq:d1nc-parameterisation} with parameters 
\begin{equation}
    (g' , \theta') = ( 4\pi / g , - g^2 \lambda / 2) \ ,
\end{equation}
while if $\theta\neq0$, this remains a generalised SNC matrix with
\begin{equation}
    (g' , \theta' , \lambda' ) = ( g \theta / 2\pi , -4\pi^2 / \theta , \lambda + 8\pi^2 / g^2 \theta) \ ,
\end{equation}
exactly as in \eqref{eq: SNC S duality}. The same transformation on the generalised D1NC moduli matrix is\footnote{Note that we have flipped the sign of $\mathsf S$ for convenience as the two are equivalent in $\operatorname{PSL}(2;\mathbb{Z})$.}
\begin{align} \nonumber
    \mathsf S \circ \mathrm{M}_{\mathrm{gD1NC}} &= \frac{g}{\sqrt{4\pi}}
    \begin{pmatrix}
        0 & 1 \\
        -1 & 0
    \end{pmatrix} 
    \begin{pmatrix}
         \theta / 2\pi &  -4\pi / \gd^2 \\
         1 & 0
    \end{pmatrix} \\
    &= \frac{g}{\sqrt{4\pi}} \begin{pmatrix}
        1 & 0 \\
        -\theta / 2\pi & 4\pi / g^2
    \end{pmatrix} \ ,
\end{align}
which is a generalised SNC matrix with
\begin{equation}
    (g' , \theta' , \lambda' ) = ( 4\pi / g , 0 , - g^2 \theta / 8 \pi^2 ) \ ,
\end{equation}
recovering the result \eqref{eq: d1nc s duality}. Similarly, the action of $\mathsf T$ on the moduli matrices are
\begin{align} \nonumber
    \mathsf T \circ \mathrm{M}_{gSNC} &= \frac{g}{\sqrt{4\pi}} \begin{pmatrix}
        1 & 1 \\
        0 & 1 
    \end{pmatrix} \begin{pmatrix}
         4\pi/ g^2 + \lambda \theta / 2\pi &  \theta /2\pi \\
         \lambda & 1
    \end{pmatrix} \\
    &= \frac{g}{\sqrt{4\pi}} \begin{pmatrix}
         4\pi/ g^2 + \lambda (\theta + 2\pi) / 2\pi &  (\theta+2\pi) /2\pi \\
         \lambda & 1
    \end{pmatrix} \ ,
\end{align}
and
\begin{align} \nonumber
    \mathsf T \circ \mathrm{M}_{\mathrm{gD1NC}} &= \frac{g}{\sqrt{4\pi}}
    \begin{pmatrix}
        1 & 1 \\
        0 & 1
    \end{pmatrix} 
    \begin{pmatrix}
         \theta / 2\pi & - 4\pi / \gd^2 \\
         1 & 0
    \end{pmatrix} \\
    &= \frac{g}{\sqrt{4\pi}} \begin{pmatrix}
         (\theta+2\pi) / 2\pi & - 4\pi / \gd^2 \\
         1 & 0
    \end{pmatrix} \ ,
\end{align}
which implements the $\theta$-angle shifts \eqref{eq: t dualities}. From these we can construct further dualities: for instance, conjugating $\mathsf T$ with $S$ gives
\begin{equation}
    \mathsf{U} = \mathsf S^{-1} \circ \mathsf T \circ \mathsf S= \begin{pmatrix}
        1 & 0 \\
        -1 & 1
    \end{pmatrix} \ .
\end{equation}
Applying this to the generalised SNC moduli matrix (where we use $\mathsf{T}$ transformations to keep $\theta\in [0,2\pi]$, with the two endpoints identified, for convenience) gives
\begin{align}
    \mathsf U \circ \mathrm{M}_{\mathrm{gSNC}} &= \frac{g}{\sqrt{4\pi}} \begin{pmatrix}
        1 & 0 \\
        -1 & 1
    \end{pmatrix} \begin{pmatrix}
         4\pi/ g^2 + \lambda \theta / 2\pi &  \theta /2\pi \\
         \lambda & 1
    \end{pmatrix} \\
    &= \frac{g}{\sqrt{4\pi}} \begin{pmatrix}
        4\pi/ g^2 + \lambda \theta / 2\pi &  \theta /2\pi \\
        \lambda (1 - \theta/2\pi) - 4\pi / g^2 & 1 - \theta / 2\pi 
    \end{pmatrix} \ .
\end{align}
If $\theta = 2\pi$ this maps us to a gD1NC theory, as one would expect from chain of dualities in the definition of $\mathsf U$. Perhaps more interestingly, when $\theta = 0$ this is an gSNC theory with 
\begin{equation}
    (g', \theta' , \lambda') = (g , 0 , \lambda - 4\pi / g^2 ) \ ,
\end{equation}
meaning $\lambda$ is periodic when $\theta = 0$. Using the form \eqref{eq: snc parameter def} of the parameters we see that this occurs when $\beta = 0$, with the new $\operatorname{SL}(2;\mathbb{R})$ parameters $\alpha' = \alpha$, $\delta' = \delta$, $\beta' = \beta = 0$, and
\begin{equation} \label{eq: gamma prime definition}
    \gamma' = \frac{\gamma \delta - 1}{\delta} \ .
\end{equation}
We shall return to this point in the next section. If we instead apply $\mathsf U$ to the gD1NC moduli, we find
\begin{align} \nonumber
    \mathsf U \circ \mathrm{M}_{\mathrm{gD1NC}} &= \frac{g}{\sqrt{4\pi}} \begin{pmatrix}
        1 & 0 \\
        -1 & 1
    \end{pmatrix} \begin{pmatrix}
         \theta / 2\pi & - 4\pi / \gd^2 \\
         1 & 0
    \end{pmatrix} \\
    &= \frac{1}{\sqrt{4\pi}} \left( \frac{4\pi}{g} \right) \begin{pmatrix}
        g^2 \theta / 8\pi^2 & -1 \\
        g^2 (1 - \theta / 2\pi) / 4\pi & 1
    \end{pmatrix} \ ,
\end{align}
giving a gSNC theory with
\begin{equation}
    (g' , \theta' , \lambda') = (4\pi / g , - 2\pi , g^2( 1 - \theta / 2\pi) / 4\pi) \ .
\end{equation}

The physically distinct moduli belong to the space of matrices $\widetilde{\mathcal M}\equiv\operatorname{PSL}(2 ; \mathbb{R})$ quotiented by the left action of $\operatorname{PSL}(2 ; \mathbb{Z})$:
\begin{equation}
    \mathcal{M} = \operatorname{PSL}(2 ; \mathbb{Z}) \setminus \widetilde{\mathcal M} \ .
\end{equation}
As $\operatorname{PSL}(2 ; \mathbb{Z})$ is not a normal subgroup of $\operatorname{PSL}(2 ; \mathbb{R})$, its right action does not generate dualities at a generic point in \(\widetilde{\mathcal M}\). It is simple to see this explicitly: for example, acting on the right of $\mathrm{M}_{\mathrm{gD1NC}}$ with $\mathsf S$ gives
\begin{equation}
    \mathrm{M}_{\mathrm{gD1NC}} \circ \mathsf S = \frac{g}{\sqrt{4\pi}} \begin{pmatrix}
        4\pi / g^2 &  \theta/ 2\pi \\
        0 & 1
    \end{pmatrix} \ ,
\end{equation}
which is a generalised SNC theory with the same coupling and $\theta$-angle; such a theory cannot obtained through a duality for generic values of $\theta$.

Although at a generic point in \(\tilde{\mathcal M}\coloneqq\operatorname{PSL}(2;\mathbb R)\) the right \(\operatorname{PSL}(2;\mathbb Z)\)-action does not generate dualities, at special points in \(\widetilde{\mathcal M}\), namely in the discrete lattice \(\Gamma\equiv\operatorname{PSL}(2;\mathbb Z)\subset\widetilde{\mathcal M}\), the \(\operatorname{PSL}(2;\mathbb Z)\)-action \emph{does} generate dualities: since for \(x\in\Gamma\) and \(g\in\operatorname{PSL}(2;\mathbb Z)_\mathrm R\), the point \(xg\in\widetilde{\mathcal M}\) is related to \(x\in\widetilde{\mathcal M}\) via the left \(\operatorname{PSL}(2;\mathbb Z)\)-action by \(xg^{-1}x^{-1}\in\operatorname{PSL}(2;\mathbb Z)_\mathrm L\):
\begin{equation}
    (xg^{-1}x^{-1})\cdot(xg) 
    = x\,.
\end{equation}
Hence, an enlarged group of S-dualities \(\operatorname{PSL}(2;\mathbb Z)_\mathrm L\times\operatorname{PSL}(2;\mathbb Z)_\mathrm R\) acts on the lattice \(\Gamma\subset\widetilde{\mathcal M}\). This phenomenon does not have a parallel in the relativistic \(\mathcal N=4\) theory, for which the group of S-dualities is never enlarged at any point in the moduli space.
Of course, although the S-duality \emph{group} is enlarged at special points \(\Gamma\subset\widetilde{\mathcal M}\), the \emph{orbit} of the S-duality group's action is not enlarged: any right \(\operatorname{PSL}(2;\mathbb{Z})\)-action on \(\Gamma\) can be converted into a left \(\operatorname{PSL}(2;\mathbb{Z})\)-action, and vice versa.
This is similar to how a simple centreless Lie group \(G\) has isometry group \(G\times G\) combining the left and right \(G\)-actions on itself, although the orbit of the left \(G\)-action on itself equals the orbit of the right \(G\)-action on itself.

\subsection{Matching the One\texorpdfstring{-}{‐}Particle Spectra} \label{sec: spectrum}

While we can derive S-duality for the Abelian theory, this cannot be done for non-Abelian gauge groups (which are what we are really interested in). We can, however, perform certain non-trivial checks of the proposal. For instance, if the $\operatorname{PSL}(2;\mathbb{Z})$ transformations really generate dualities of the space of theories then the limits constructed from \eqref{eq:transformed-sugra-solution} with all parameters integers must be dynamically equivalent. In particular, their one-particle spectra should match. The only difference between each theory should be the charges assigned to each state, with the charge mapping following that predicted by S-duality.

Let us start by considering the generalised D1NC theory, working with no $\theta$-term for now. As the moduli space dynamics of BPS monopoles is well-studied, we shall be brief, reviewing the key aspects for our purposes.\footnote{For more detailed expositions, see \cite{Tong:2005un, Weinberg:2006rq}.} For definiteness, we shall consider the gauge group $G= \operatorname SU(2)$. As we are interested in transformations with integer coefficients we shall work with the action \eqref{eq: d1nc pre normalisation} and take $\gamma = -\beta = 1$. The Lagrange multiplier $G_{ij}$ imposes the BPS monopole equation
\begin{equation} \label{eq: bps monopole eq}
    F_{ij} = \varepsilon_{ijk} D_k X \ .
\end{equation}
The one-monopole states are given by the Prasad--Sommerfield solution \cite{Prasad:1975kr} of \eqref{eq: bps monopole eq}, which in our conventions is taken to be
\begin{subequations} \label{eq: ps solution}
\begin{align}
    X &= \frac{\hat{r}^i \sigma^i}{2} \frac{1 - vr \coth(vr)}{r} \ , \\
    A_i &= \frac{ \varepsilon_{ijk} \hat{r}^j \sigma^k}{2} \frac{1 - vr \csch(vr)}{r} \ ,
\end{align}
\end{subequations}
along with its moduli. These are a shift in the monopole position (taken to be the origin of $\mathbb{R}^3$ in \eqref{eq: ps solution}), which we denote by $x_0^i$, and the $U(1)$ gauge transformation of $A_i$ generated by $X$, which we denote as $\chi$. The normalisation of $v$ is taken such that
\begin{equation} \label{eq: surface integral of X}
    \int_{\mathbb S_{\infty}^2} \frac{\mathrm d S_i}{2\pi} \tr ( X D_i X ) = v \ ;
\end{equation}
whenever we have a surface integral from here onwards it will always be taken on the asymptotic 2-sphere \(\mathbb S_{\infty}^2\), so we shall drop the explicit appearance of $\mathbb S_{\infty}^2$ to ease notation.

Since \eqref{eq: bps monopole eq} is time-independent, the full theory is given by the Manton moduli-space approximation \cite{Manton:1981mp} for monopole scattering; time evolution enters the theory by promoting the moduli to be functions of $t$. On the moduli space the D1NC action reduces to
\begin{equation}
    S_{\mathrm{D1NC}} = \int \frac{\mathrm dt\,\mathrm d^3x}{4\pi} \tr\bigg( F_{0i} F_{0i} + D_0X D_0X 
    \bigg) \ .
\end{equation}
Using the infinitesimal transformation of \eqref{eq: ps solution} under changes of the moduli to find the BPS monopole equation's zero-modes\footnote{We must also gauge-fix $A_0$ to ensure that the zero modes are orthogonal; see the aforementioned reviews of the subject for a full account of this.}, the fields entering the action are
\begin{subequations}
\begin{align} \label{eq: f0i zero modes}
    F_{0i} &= - \dot{x}_0^j F_{ij} + \frac{ \dot{\chi}}{v} D_i X \ , \\
    D_0 X &= - \dot{x}_0^i D_i X \ .
\end{align}
\end{subequations}
The prefactor in the $\dot{\chi}$ term is required to properly normalise the $U(1)$ generator proportional to $X$, ensuring that $\chi \sim \chi + 2\pi$. From these and \eqref{eq: surface integral of X} it is easy to see that the D1NC action is
\begin{equation} \label{eq: moduli space action}
    S_{\mathrm{D1NC}} = \int\mathrm{dt} \, \bigg( \frac{v \dot{x}_0^2}{2} + \frac{\dot{\chi}^2}{2 v}  \bigg) \ .
\end{equation}
The eigenfunctions of the Hamiltonian one obtains from \eqref{eq: moduli space action} are
\begin{equation}
    \psi_{p,k}(x,\chi) = \exp\mleft(\mathrm i (p \cdot x + k \chi)\mright) \ ,
\end{equation}
where $p\in \mathbb{R}^3$ and $k\in\mathbb{Z}$, and the corresponding eigenvalues (dispersion relations) are
\begin{equation} \label{eq: D1nc spectrum}
    E_k(p) = \frac{p^2}{2v} + \frac{v k^2}{2} \ .
\end{equation}
We have focused solely on the bosonic sector of the theory here. It is straightforward to see that the fermionic sector of the supersymmetric extension of \eqref{eq: moduli space action} has vanishing Hamiltonian. The effect of the fermions is therefore to generate a supermultiplet of particle states of different four-dimensional spin but equal energy; for the purposes of matching energies we can therefore focus only on the bosonic sector for the rest of this section.

The magnetic charge of for the $\operatorname U(1)$ subgroup of $\operatorname{SU}(2)$ generated by a Lie-algebra element $\sigma\in\mathfrak{su}(2)$ is given by
\begin{equation}
    P[\sigma] = \int \frac{\mathrm d S_i}{4\pi} \, \varepsilon_{ijk} \tr( \sigma F_{jk} ) \ .
\end{equation}
Taking $\sigma = v^{-1} X$, this evaluates on a BPS monopole to $P[\sigma] = 1$. Similarly, the electric charge of the states with respect to the same $U(1)$ subgroup is given by
\begin{equation}
    Q[\sigma] = \int \frac{\mathrm d S_i}{2\pi} \, \tr( \sigma F_{0i} ) \ ,
\end{equation}
which evaluates on \eqref{eq: f0i zero modes} to
\begin{equation}
    Q = \frac{\dot{\chi}}{v} \equiv \pi_{\chi} \ ,
\end{equation}
the momentum conjugate to $\chi$. Using the form of the eigenfunctions above we see that
\begin{equation}
    \hat{Q} \psi_{p,k} = k \psi_{p,k}
\end{equation}
after quantising. What about the $\theta$-term? In general, it has two effects. Evaluating it on the solution using \eqref{eq: f0i zero modes} and the BPS monopole equation, we see that
\begin{align} \nonumber
    S_{\theta} &=  \int \frac{\mathrm dt\,\mathrm d^3x}{2\pi}\, \frac{\theta \dot{\chi}}{2\pi v} \tr( D_i X D_i X) \\ 
    &= \int \mathrm dt \,\frac{\theta \dot{\chi}}{2\pi} \ .
\end{align}
This shifts the momentum conjugate to $\chi$ to
\begin{equation}
    \pi_{\chi} = \frac{\dot{\chi}}{v} + \frac{\theta}{2\pi} \ ,
\end{equation}
taking $k \mapsto k + \frac{\theta}{2\pi}$ in \eqref{eq: D1nc spectrum}. However, in our case $\theta = 2\pi \alpha$, with $\alpha$ an integer, so there is no change in the spectrum. The other role of the $\theta$-term is to alter the electric charge \cite{Witten:1979ey}, taking it to
\begin{equation}
    \Tilde{Q}[\sigma] = \int \frac{\mathrm d S_i}{2\pi} \tr \sigma \left( F_{0i} + \frac{\alpha}{2} \varepsilon_{ijk} F_{jk} \right) \ .
\end{equation}
On the one-monopole solution, this is
\begin{equation}
    \Tilde{Q}[\sigma] = Q[\sigma] + \alpha \ ,
\end{equation}
meaning the generalised D1NC theory's one particle spectrum is given by \eqref{eq: D1nc spectrum} with the charges
\begin{equation}
    (\Tilde{Q}, P) = (\alpha + k, 1) \ .
\end{equation}

Let us now turn our attention to the generalised SNC theory, which we shall work with in the form \eqref{eq: gsnc action pre-rescaling} to make the matching easy to see. The equations of motion of the gauge sector (setting $Y^A=0$) are
\begin{subequations}
\begin{align} \label{eq: snc eq 1}
    0 &= \mathrm i [X,D_0 X ] - \delta^{-1} D_i D_i X \ , \\
    0 &=  D_0 D_i X + \mathrm i [X , F_{0i} ] + \delta^{-1} D_j F_{ij} - \mathrm i \gamma^2 \delta^{-1} [X , D_i X] \ , \\
    0 &= D_0^2 X + \delta^{-1} D_i F_{0i} - \gamma^2 \delta^{-2} D_i D_i X \ . 
\end{align}
\end{subequations}
These are solved by the static equations
\begin{subequations} \label{eq: dyon solution of gsnc}
\begin{align}
    F_{ij} &= \Tilde{\gamma} \varepsilon_{ijk} D_k X \ , \\
    A_0 &=  \frac{\Tilde{\gamma}^2 - \gamma^2 }{2\delta} X \ ,
\end{align}
\end{subequations}
with $\partial_0 X = 0$. The solutions can be found by scaling the Prasad--Sommerfield solution \eqref{eq: ps solution}, giving
\begin{subequations} \label{eq: scaled ps solution}
\begin{align}
    X &= \frac{\hat{r}^i \sigma^i}{2} \frac{\Tilde{\gamma} - \frac{ v r}{\Tilde{\gamma}} \coth\left(\frac{ v r }{\Tilde{\gamma}^2}\right)}{r} \ , \\
    A_i &= \frac{ \varepsilon_{ijk} \hat{r}^j \sigma^k}{2} \frac{1 - \frac{ vr}{\Tilde{\gamma}^2} \csch\left(\frac{ v r}{\Tilde{\gamma}^2}\right)}{r} \ ;
\end{align}
\end{subequations}
we have normalised $v$ to ensure that \eqref{eq: surface integral of X} remains true.

The energy of the gauge sector with the conventions \eqref{eq: gsnc action pre-rescaling} is
\begin{equation} \label{eq: gsnc energy}
    E = \int \frac{\mathrm d^3 x}{4\pi} \tr \bigg( (D_0 X)^2 + \frac{1}{2\delta^2} (F_{ij} - \gamma \varepsilon_{ijk} D_k X)^2  \bigg) \ ,
\end{equation}
which evaluates on the solution \eqref{eq: dyon solution of gsnc} to
\begin{align} \nonumber
    E &= \frac{\Delta\gamma^2}{2 \delta^2} \int \frac{\mathrm dS_i}{2\pi} \tr ( X D_i X) \\
    &= \frac{\Delta\gamma^2 v}{2 \delta^2} \ ,
\end{align}
after defining $\Delta\gamma = \Tilde{\gamma} - \gamma $ and using \eqref{eq: surface integral of X}. The magnetic charge for the $\operatorname U(1)$ subgroup of $\operatorname{SU}(2)$ associated with a gauge transformation $\sigma$ is
\begin{equation}
    P[\sigma] =  \int \frac{\mathrm dS_i}{4\pi} \varepsilon_{ijk} \tr( \sigma  F_{jk} ) \ ,
\end{equation}
while the corresponding electric charge computed via the usual Noether method is
\begin{equation}
    Q[\sigma] = \int \frac{\mathrm dS_i}{2\pi} \tr \sigma \bigg( \frac{1}{\delta} D_i X + \frac{\beta}{2\delta} \varepsilon_{ijk} F_{jk}  \bigg) \ .
\end{equation}
Making the choice $\sigma = v^{-1} X$, substituting \eqref{eq: dyon solution of gsnc} into the charges, and evaluating using \eqref{eq: surface integral of X} then gives
\begin{align} \nonumber
    Q[\hat{\phi}] &= \frac{1 + \beta (\gamma + \Delta \gamma)}{\delta v} \int \frac{dS_i}{2\pi} \tr ( X D_i X ) \\
    &= \left( \alpha + \frac{\beta \Delta\gamma}{\delta} \right) \ ,
\end{align}
using $\alpha \delta - \beta \gamma = 1$, and
\begin{align}
    P[\hat{\phi}] = \gamma + \Delta \gamma \ .
\end{align}

Since $\beta$ and $\delta$ are coprime integers, asking for quantisation of the charges forces us to take
\begin{equation}
    \Delta \gamma = k \delta
\end{equation}
in terms of $k\in\mathbb{Z}$. We therefore find a tower of static states labelled by an integer $k$ with charges
\begin{equation} \label{eq: charge tower}
    (Q_k[\hat{\phi}] , P_k[\hat{\phi}] ) =  (\alpha , \gamma) + k (\beta , \delta)
\end{equation}
and energies
\begin{equation} \label{eq: snc rest energy}
    E_k = \frac{v k^2}{2} \ .
\end{equation}
The static one-particle states therefore match those of the D1NC theory; furthermore, the mapping between the charges is exactly as one would expect from the $\operatorname{SL}(2;\mathbb{Z})$ transformations. As $k$ runs over all integers, if we redefine it by an integer shift, we find an equivalent spectrum with the shifted charges ground state charges
\begin{equation}
    (\alpha' , \gamma ' ) = (\alpha - \beta \Delta k , \gamma - \delta \Delta k ) \ .
\end{equation}
If we consider the special case $\beta = 0$, for which $\alpha = \delta = 1$, then we find that the theory is invariant under taking $\gamma \mapsto \gamma - 1$, exactly as predicted from the duality transformation in \eqref{eq: gamma prime definition} with $\operatorname{SL}(2;\mathbb{Z})$ parameters. 

One may hope that we can find non-static solutions as in the D1NC theory by giving time dependence to the moduli of the one-particle solution. We shall see in the next section that the generalised SNC action (and hence its equations of motion) is invariant under the coordinate transformations
\begin{subequations}
\begin{align}
    \hat{t} &= t \ , \\
    \hat{x}^i &= x^i + \xi^i(t) \ ,
\end{align}
\end{subequations}
for any $\xi^i(t)$, provided we take the field transformations
\begin{subequations}
\begin{align}
    \hat{A}_0(\hat{t} , \hat{x}) &= A_0(t,x) - \dot{\xi}^i A_i(t,x) \ , \\
    \hat{A}_i(\hat{t}, \hat{x}) &= A_i(t,x) + \delta \dot{\xi}^i X (t,x) \ , \\
    \hat{X}(\hat{t}, \hat{x}^i) &= X(t,x) \ ,
\end{align}
\end{subequations}
at linear order in $\xi^i$ for the fields relevant for our current purposes\footnote{This differs slightly from \eqref{eq: SNC boost X 2} due to differing normalisation of $X$ in this section.}. However, we shall see there that this is only a symmetry of the boundary conditions of our fields if we restrict $\xi^i$ to be at most linear in $t$. If we let $\{ \bar{A}_i(x) , \bar{X}(x)\}$ denote the static solution \eqref{eq: scaled ps solution}, we can therefore use this to motivate the solution\footnote{We thank the anonymous referee for pointing out the existence of this solution.}
\begin{subequations}
\begin{align}
    X(t,x) &= \bar{X}(x - u t) \ , \\
    A_i(t,x) &= \bar{A}_i(x - u t) + \delta\, u^i \bar{X}(x - u t) \ , \\
    A_0(t,x) &= \left(\frac{\Tilde{\gamma}^2 - \gamma^2}{2\delta} - \frac{u^2 \delta}{2}\right) \bar{X}(x - u t) - u^i \bar{A}_i(x - u t) \ ,
\end{align}
\end{subequations}
obeying the same equations and boundary conditions travelling at a uniform velocity $u^i$. It is straightforward to compute the energy of the solution using \eqref{eq: gsnc energy}; the cross-terms between the original static solution and those induced by $u^i$ cancel, leaving
\begin{equation}
    E_k = \frac{v u^2}{2} + \frac{v k^2}{2} \ .
\end{equation}
This exactly matches the spectrum \eqref{eq: D1nc spectrum} found in the D1NC theory if we associate
\begin{equation}
    p^i = v u^i \ ,
\end{equation}
as one would expect for a non-relativistic particle of mass $v$.

It is interesting to ask what happens if we consider a solution with no non-trivial scalar contribution (in other words, constant $X$ and vanishing $A_i$). For definiteness let us take $X = v\sigma$ (with $\sigma$ a semisimple element of $\mathfrak{su}(2)$) and combine the algebra's other two generators into the pair $\sigma^{\pm}$ defined by
\begin{subequations}
\begin{align}
    [\sigma, \sigma^{\pm}] &= \pm \sigma^{\pm} \ , \\
    \tr( \sigma^+ \sigma^-) &= 1 \ .
\end{align}
\end{subequations}
We can then consider the theory's perturbative excitations, which we shall illustrate using a single component of $Y^A$ that we shall refer to as $Y$. Truncating the action of $Y$ to its quadratic terms gives
\begin{equation}
    S_Y = \int \frac{ \mathrm dt\,\mathrm d^3 x}{8\pi} \tr \bigg( 2\mathrm i v \delta [\sigma, Y] \partial_0 Y - (\partial_i Y)^2 + \gamma^2 v^2 [\sigma, Y]^2 \bigg) \ .
\end{equation}
The dynamically interesting part of $Y$ comes from the $\sigma^{\pm}$ components; expanding the field as
\begin{equation}
    Y = \sqrt{\frac{4\pi }{v \delta} } \bigg( \bar{\omega} \sigma^+ + \omega \sigma^- \bigg) \ ,
\end{equation}
gives us the canonically normalised action
\begin{equation}
    S_Y = \int\mathrm dt\,\mathrm d^3 x \bigg(i \bar{\omega} \partial_0 \omega - \frac{1}{2v \delta} \partial_i \bar{\omega} \partial_i \omega - \frac{\gamma^2 v}{\delta} \bar{\omega} \omega \bigg) \ .
\end{equation}
The equation of motion of $\bar{\omega}$ is
\begin{equation}
    \mathrm i \partial_0 \omega + \frac{1}{2v \delta } \partial_i \partial_i \omega - \frac{\gamma^2 v}{\delta} \omega = 0 \ ,
\end{equation}
from which we can read off the energy
\begin{equation} \label{eq: gsnc spectrum}
    E = \frac{p^2}{2v \delta} + \frac{\gamma^2 v}{2 \delta} \ .
\end{equation}
The energy gap matches \eqref{eq: snc rest energy} if we take
\begin{equation}
    \gamma^2 = k^2 \ , \quad
    \delta = 1 \ .
\end{equation}
With $\gamma = - k$, this is what one would expect from the charge assignment \eqref{eq: charge tower}, since the magnetic charge then vanishes as it should for a perturbative excitation of the QFT. It would be interesting to take this further and quantise the field theory in the static background \eqref{eq: scaled ps solution} to see whether we can find further matching with the gSNC semiclassical spectrum \eqref{eq: gsnc spectrum}.

\section{Matching Symmetries}\label{sec:symmetry}

\subsection{Spacetime Galilean Symmetries}

If the theories \eqref{eq: gen SNC action} and \eqref{eq: gD1 action} are to map into each other under $\operatorname{SL}(2;\mathbb{Z})$ transformations, their symmetries must be equivalent. These were computed for the undeformed D1NC theory in \cite{Lambert:2024yjk, Lambert:2024ncn}, with the bosonic symmetries of the  undeformed SNC theory found in \cite{Bagchi:2022twx, Fontanella:2024hgv}\footnote{In both cases the transformations which appear in this work are slightly altered from these to account for differences in conventions.}. In each case the same symmetries appear on both sides, and it will be instructive to show that this is preserved by the generalisations of both. 

Let us start with the (bosonic) spacetime symmetries, beginning with those that act on $t$. The actions are invariant up to total derivatives for the infinitesimal temporal conformal transformations
\begin{subequations} \label{eq: conformal transformations}
\begin{align}
    \hat{t} &= t + a + b t + ct^2 \ , \\
    \hat{x}^i &= \left(1 + b + 2 c t\right) x^i \ ,
\end{align}
\end{subequations}
combined with the field transformations
\begin{subequations}
\begin{align}
    \hat{A}_0(\hat{t}, \hat{x}) &= \left(1 - b - 2ct \right) A_0(t,x) - 2 c x^i A_i(t,x) \ , \\
    \hat{A}_i(\hat{t}, \hat{x}) &= \left(1 - b - 2ct \right) A_i(t,x) + 2 c x^i X(t,x) \ , \\
    \hat{X}(\hat{t}, \hat{x}) &= \left( 1 - b - 2c t \right) X(t,x) \ , \\
    \hat{Y}^M(\hat{t}, \hat{x}) &= \left( 1 - b - 2c t \right) Y^M(t,x) \ ,
\end{align}
\end{subequations}
in the SNC theory, and
\begin{subequations}
\begin{align}
    \hat{A}_0(\hat{t}, \hat{x}) &= \left(1 - b - 2ct \right) A_0(t,x) - 2 c x^i A_i(t,x) \ , \\
    \hat{A}_i(\hat{t}, \hat{x}) &= \left(1 - b - 2ct \right) A_i(t,x) \ , \\
    \hat{X}(\hat{t}, \hat{x}) &= \left( 1 - b - 2c t \right) X(t,x) \ , \\
    \hat{Y}^M(\hat{t}, \hat{x}) &= \left( 1 - b - 2c t \right) Y^M(t,x) \ , \\ \nonumber
    \hat{G}_{ij}(\hat{t}, \hat{x}) &= \left( 1 - 2b - 4c t \right)G_{ij}(t,x) - 4ct  x_{[i} F_{|0|j]}(t,x) \\
    &\qquad+ 2 c t \varepsilon_{ijk} x^k D_0 X(t,x) \ ,
\end{align}
\end{subequations}
for the D1NC theory. It is straightforward to see that this is also the case once the $\theta$-term is included; the field strength components have the transformations
\begin{subequations}
\begin{align}
    \hat{F}_{0i}(\hat{t}, \hat{x}) &= (1 - 2b - 4 c t) F_{0i}(t,x) + 2 c x^j F_{ij}(t,x) + 2  \sigma c x^i D_0 X(t,x) \ , \\ \label{eq: F ij transformation}
    \hat{F}_{ij}(\hat{t}, \hat{x}) &= (1 - 2b - 4 c t) F_{ij}(t,x)  - 4 \sigma c x_{[i} D_{j]} X(t,x) \ , 
\end{align}
\end{subequations}
where $\sigma= 0$ for the generalised D1NC limit and 1 for the generalised SNC limit, meaning the term's transformation is
\begin{align} 
    \deltaup S_{\theta} = \frac{\theta c}{(2\pi)^2} \tr \int \mathrm dt\,\mathrm d^3 x \, \bigg(& \varepsilon_{ijk} F_{ij} F_{kl} x^l - \sigma  \varepsilon_{ijk} x^i F_{jk} D_0 X + 2 \sigma \varepsilon_{ijk} x^i D_j X F_{0k} \bigg) \ .
\end{align}
Using the identities
\begin{align} \label{eq: identities}
    \varepsilon_{ijk} \tr \, F_{ij} F_{kl} &= 0 \ , &
    D_0 F_{ij} - 2 D_{[i} F_{|0|j]} &= 0 \ ,
\end{align}
we see that $\deltaup S_{\theta} = 0$ up to $X$-dependent boundary terms. This is sufficient for the generalised D1NC theory, but for the generalised SNC theory we must also consider the deformation of the spatial field strength's quadratic term\footnote{The transformation of the $[X, Y^A]^2$ term is homogeneous and possesses the correct scaling dimension, so it is automatically invariant.}. The new contribution from the inhomogeneous term in \eqref{eq: F ij transformation} vanishes due to symmetry, meaning the transformations are symmetries of both the generalised D1NC and SNC actions. However, the interesting dynamics of both theories take place on the Coulomb branch, with a VEV given to $X$; it is clear that the scaling transformations in \eqref{eq: conformal transformations} are broken, leaving us with only time-translational invariance.

Next we have the transformations that leave $t$ invariant. At the level of the action, these are infinitesimal spatial translations with arbitrary time-dependence,
\begin{equation} \label{eq: time dependent translations}
    \hat{x}^i = x^i + \xi^i(t) \ ,
\end{equation}
and spatial rotations. It is clear that the deformations preserve rotational symmetry and so we will focus here on the translations, where the non-trivial SNC field transformations are
\begin{subequations}
\begin{align}
    \hat{A}_0(\hat{t}, \hat{x}) &= A_0(t,x) - \dot{\xi}^i A_i(t,x) \ , \\ \label{eq: SNC boost X 2}
    \hat{A}_i(\hat{t}, \hat{x}) &= A_i(t,x) + \dot{\xi}^i X(t,x) \ ,
\end{align}
\end{subequations}
and the non-trivial D1NC transformations are
\begin{subequations}
\begin{align}
    \hat{A}_0(\hat{t}, \hat{x}) &= A_0(t,x) - \dot{\xi}^i A_i(t,x) \ , \\
    \hat{G}_{ij}(\hat{t}, \hat{x}) &= G_{ij}(t,x) + 2 F_{0[i} \dot{\xi}_{j]} + \varepsilon_{ijk} \dot{\xi}^k D_0 X + \varepsilon_{ijk} \ddot{\xi}^k X \ .
\end{align}
\end{subequations}
The action of this on the $\theta$-term is nearly identical to that of the conformal transformations, with the transformations
\begin{subequations}
\begin{align}
    \hat{F}_{0i}(\hat{t}, \hat{x}) &= F_{0i}(t,x) + \dot{\xi}^j F_{ij}(t,x) + \sigma D_0 \big( \dot{\xi}^i X \big)(t,x) \ , \\
    \hat{F}_{ij}(\hat{t}, \hat{x}) &= F_{ij}(t,x) - 2 \sigma \dot{\xi}_{[i} D_{j]}X(t,x) \ ,
\end{align}
\end{subequations}
leading to
\begin{align} \nonumber
    \deltaup S_{\theta} = \frac{\theta}{2(2\pi)^2} \tr \int\mathrm dt\,\mathrm d^3 x \, \bigg(& \varepsilon_{ijk} F_{ij} F_{kl} \dot{\xi}^l - \sigma  \varepsilon_{ijk}  F_{jk} D_0 \big( \dot{\xi}^i X) \\
    &+ 2 \sigma \varepsilon_{ijk} \dot{\xi}^i D_j X F_{0k} \bigg) \ ,
\end{align}
and so we find invariance of the action through the use of the identities \eqref{eq: identities}. It can be similarly seen that the deformation of the spatial field strength's quadratic term remains invariant under the inhomogeneous term in the SNC transformations, meaning that both the generalised D1NC and SNC actions are invariant under \eqref{eq: time dependent translations}. As we are really interested in the physics on the Coulomb branch, we must account for the VEV of $X$ which fixes its asymptotic behaviour as 
\begin{equation}
    X \to \langle X \rangle \ .
\end{equation}
We then see that the transformations act asymptotically on $G_{ij}$ for the D1NC-type theories,
\begin{equation}
    \deltaup G_{ij} \to \varepsilon_{ijk} \ddot{\xi}^k \langle X \rangle \ ,
\end{equation}
and $F_{0i}$ for the SNC-type theories,
\begin{equation}
    \deltaup F_{0i} \to \ddot{\xi}^i \langle X \rangle \ ,
\end{equation}
altering the theory's boundary conditions in both cases. We must therefore restrict to $\ddot{\xi}=0$, meaning the Coulomb branch theory is only invariant under translations and Galilean boosts. To summarise, the spacetime symmetry algebra of the generalised D1NC and SNC actions is generated by the temporal conformal transformation \eqref{eq: conformal transformations}, spatial rotations, and time-dependent translations \eqref{eq: time dependent translations}, forming a conformal extension of the Milne algebra \cite{Duval:1993pe}. However, if we are working on the Coulomb branch this is broken to the more familiar three-dimensional Galilean group formed from spacetime translations, spatial rotations, and Galilean boosts.

\subsection{Electric and Magnetic One\texorpdfstring{-}{‐}Form Symmetries}

While we have seen that the spacetime symmetries of the SNC and D1NC theories match, we should also check that this holds their one-form symmetries. Let us briefly recall how this works in the relativistic theory. Any four-dimensional Yang--Mills theory with gauge group \(G\) with adjoint matter (such as the maximally supersymmetric theory) enjoys one-form electric and magnetic symmetries, where the electric symmetry is valued in the centre subgroup \(\operatorname Z(G)\) and the magnetic symmetry is valued in the Pontryagin dual group to the fundamental group, \(\widehat{\operatorname\pi_1(G)}\). For instance, for Maxwell theory, where \(G=\operatorname U(1)\), the electric symmetry is valued in \(\operatorname Z(\operatorname U(1))=\operatorname U(1)\), and the magnetic symmetry in \(\widehat{\operatorname\pi_1(\operatorname U(1))}=\hat{\mathbb Z}=\operatorname U(1)\); for the simply connected gauge group \(G=\operatorname{SU}(N)\), the electric symmetry is valued in \(\operatorname Z(\operatorname{SU}(N))=\mathbb Z_N\), and the magnetic symmetry in \(\widehat{\pi_1(\operatorname{SU}(N))}=\hat1=1\); for the centreless gauge group \(G=\operatorname{PSU}(N)=\operatorname{SU}(N)/\operatorname Z(\operatorname{SU}(N))\), the electric symmetry is valued in \(\operatorname Z(\operatorname{PSU}(N))=1\), and the magnetic symmetry in \(\widehat{\pi_1(\operatorname{PSU}(N))}=\widehat{\mathbb Z_N}=\mathbb Z_N\).

In four-dimensional Maxwell theory without matter, the equations of motion (Maxwell's equations)
\begin{align}
    \mathrm dF &= 0,&\mathrm d\star F&=0
\end{align}
imply that \(F\) is a Noether current for a \(\operatorname U(1)\)-valued magnetic one-form symmetry and that \(\star F\) is a Noether current for a \(\operatorname U(1)\)-valued electric one-form symmetry. In the presence of electric currents \(J_\mathrm e\) or magnetic currents \(J_\mathrm m\), these equations are modified to
\begin{align}
    \mathrm dF &= J_\mathrm m,&\mathrm d\star F&=J_\mathrm e,
\end{align}
signalling the breaking of these one-form symmetries. If electric or magnetic charge is quantised, then parts of the one-form symmetries survive --- for instance, if electric charge comes in multiples of \(k\) (because there are only charge \(k\) matter fields), then a \(\mathbb Z_k\) electric one-form symmetry survives.

Let us now turn our attention to the modification of Maxwell's equations in the Abelian generalised D1NC theory and the Abelian generalised SNC theory (a.k.a.\ Galilean electromagnetism). In fact, the topological terms are total derivatives and do not change the equations of motion, so it suffices to look at the usual D1NC and SNC theories, i.e.\ with \(\theta_\mathrm{D1NC}=0\) and \(\theta_\mathrm{SNC}=\lambda=0\). 

\subsubsection{One\texorpdfstring{-}{‐}Form Symmetries in Abelian D1NC and SNC Theories}

\paragraph{Galilean Electromagnetism}
Let us consider the gauge sector of the Abelian SNC theory, which (as discussed in section \ref{sec:duality}) is Galilean electromagnetism,
\begin{equation}
    S_\mathrm{SNC}
    =\int\frac{\mathrm dt\,\mathrm d^3x}{2 g_\mathrm{SNC}^2}\,
    \left(
        (\partial_0X)^2
        + 2 \partial_iX F_{0i}
        -\frac12 F^{ij}F_{ij}
    \right).
\end{equation}
The equations of motion of the theory can be combined into
\begin{subequations}
\begin{align} \label{eq: other abelian snc equation}
    0 &= \partial_{\mu} \partial_0 X + \partial_j F_{\mu j} \ , \\ \label{eq: Laplacian of X}
    0 &= \partial_i \partial_i X \ .
\end{align}
\end{subequations}
Now, the Bianchi identity \(\mathrm dF_2=0\) continues to hold, but in general \(\mathrm d\star F_2\ne0\), that is, the electric one-form symmetry from the relativistic theory breaks while the magnetic one-form symmetry survives. This is consistent with the fact that this is the theory obtained by putting on an electric charge, which breaks the electric one-form symmetry. (Since the effective electric charge in \eqref{eq: other abelian snc equation} is not quantised, no discrete remnant of electric one-form symmetry survives.)

However, the following modified electric symmetry survives: consider the two-form
\begin{align} \nonumber
    \tilde F_2 &= -\star\left(\mathrm dt\wedge\mathrm dX+\frac12F_{ij}\,\mathrm dx^i\wedge \mathrm dx^j\right) \\
    &= \frac12\varepsilon_{ijk} \partial_i X\,\mathrm dx^j \wedge\mathrm dx^k - \frac{1}{2} \varepsilon_{ijk} F_{jk}\,\mathrm dt \wedge\mathrm dx^i \ .
\end{align}
Half of its components are those of \(\star F\), but the other half is replaced with components of \(\mathrm dX\). Then \eqref{eq: other abelian snc equation} and \eqref{eq: Laplacian of X} imply that
\begin{equation} 
    \mathrm d\tilde F_2 = \partial_i \partial_i X\,\mathrm dx^1 \wedge\mathrm dx^2 \wedge\mathrm dx^3  + \frac12\varepsilon_{ijk} ( \partial_0 \partial_i X + \partial_l F_{il} )\mathrm dt \wedge\mathrm dx^j \wedge\mathrm dx^k = 0 \ .
\end{equation}
Together with the Bianchi identity \(\mathrm dF_2\) where \(F_2 = \mathrm dA\) is the usual two-form field stregnth, we therefore obtain two one-form conserved charges
\begin{align} 
    Q^{(\mathrm{SNC})}[\Sigma_2] &= \int_{\Sigma_2} \tilde F_2\ , &
    P^{(\mathrm{SNC})}[\Sigma_2] &= \int_{\Sigma_2} F_2 \ .
\end{align}
Specialised to the case where \(\Sigma_2\) is purely spatial, these reduce (neglecting overall factors) to
\begin{subequations}
\begin{align} \label{eq: electric charge}
    Q^{(\mathrm{SNC})}[\Sigma_2] &= \int_{\Sigma_2} \mathrm dS_i \, \partial_i X \ , \\ \label{eq: magnetic charge}
    P^{(\mathrm{SNC})}[\Sigma_2] &= \int_{\Sigma_2} \mathrm dS_i \, \varepsilon_{ijk} F_{jk} \ ,
\end{align}
\end{subequations}
If we include a Wilson line in the theory, which we write as
\begin{equation} 
    S_{j_0 A_0} = q \int\mathrm dt \, A_0(t, \gamma(t))\equiv \int \frac{\mathrm dt\,\mathrm d^3 x}{ 2\gsnc^2} \, A_0 j_0 \ ,
\end{equation}
with $j_0(x,t) = 2q \gsnc^2  \deltaup^{(3)}(x - \gamma(t))$ for a curve $\gamma^i(t)$, then we turn on a source for \eqref{eq: Laplacian of X},
\begin{equation}
    - \partial_i \partial_i X = q \gsnc^2 \deltaup^{(3)}(x - \gamma(t)) \ .
\end{equation}
As one would expect, the Wilson line is therefore the object charged under the symmetry generated by \eqref{eq: electric charge}. As the magnetic charge \eqref{eq: magnetic charge} is unchanged from the Lorentzian theory, its charged objects are 't Hooft lines that create Dirac monopole configurations.

\paragraph{D1NC Theory} 
In the D1NC theory, the terms in the gauge sector are
\begin{equation}
    S_{\mathrm{D1}} = \int\frac{\mathrm dt\,\mathrm d^3x}{2g_\mathrm{D1NC}^2}\,\left((F_{0i})^2 + (\partial_0 X)^2 - G_{ij}(F_{ij}-\varepsilon_{ijk}\partial_kX) \right).
\end{equation}
This has the equations of motion
\begin{subequations}
\begin{align} \label{eq: abelian constraint}
    0 &= F_{ij} - \varepsilon_{ijk} \partial_k X \ , \\ \label{eq: gauss law d1}
    0 &= \partial_i F_{0i} \ , \\ \label{eq: time deriv of F abelian d1}
    0 &=  \partial_0 F_{0i} + \partial_j G_{ij} \ , \\
    0 &=  2\partial_0^2 X + \varepsilon_{ijk} \partial_k G_{ij} \ .
\end{align}
\end{subequations}
The Bianchi identity \(\mathrm dF_2=0\) continues to hold, so the magnetic one-form symmetry in the relativistic theory survives the non-relativistic limit unscathed. However, the electric one-form symmetry in the relativistic symmetry breaks since \(\mathrm d\star F_2\ne0\).
However, if one defines the modified electric current
\begin{align} \nonumber
    \tilde F_2 &\equiv
    -\star\left(F_{0i}\,\mathrm dt\wedge\mathrm dx^i + \frac12 G_{ij}\,\mathrm dx^i \wedge\mathrm dx^j\right) \\
    &= \frac{1}{2} \varepsilon_{ijk} F_{0i} dx^j \wedge dx^k - \frac{1}{2} \varepsilon_{ijk} G_{jk} dt \wedge dx^i \ ,
\end{align}
then the equations of motion \eqref{eq: gauss law d1} and \eqref{eq: time deriv of F abelian d1} imply that \(\mathrm d\tilde F_2 = 0\).

We therefore obtain the electric and magnetic one-form conserved charges
\begin{align} 
    Q^{(\mathrm{D1})}[\Sigma_2] &= \int_{\Sigma_2} \tilde F_2\ , &
    P^{(\mathrm{D1})}[\Sigma_2] &= \int_{\Sigma_2} F_2 \ .
\end{align}
When the surface \(\Sigma_2\) is purely spatial, these reduce to
\begin{subequations}
\begin{align} \label{eq: electric charge d1}
    Q^{(\mathrm{D1})}[\Sigma_2] &= \int_{\Sigma_2} \mathrm dS_i \, F_{0i} \ , \\ \label{eq: magnetic charge d1}
    P^{(\mathrm{D1})}[\Sigma_2] &= \int_{\Sigma_2} \mathrm dS_i \, \partial_i X \ ,
\end{align}
\end{subequations}
where we have used \eqref{eq: abelian constraint}.
The configurations for which $X$ is singular are then Dirac monopoles, with the insertion of the singular boundary condition associated with the insertion of an 't Hooft line. As with Lorentzian Maxwell theory, including a timelike Wilson line gives an electric source in \eqref{eq: gauss law d1} which is measured by \eqref{eq: electric charge d1}. The two Abelian theories therefore possess the same one-form symmetries, with a clear matching between them.

\subsubsection{One\texorpdfstring{-}{‐}Form Symmetries in non-Abelian D1NC and SNC Theories}
In a non-Abelian gauge theory, generalised global symmetries do not break as one passes to the Coulomb branch where part of the gauge symmetry is Higgsed away. Thus, in the present case, the one-form symmetries for non-Abelian case (corresponding to a stack of coincident D3-branes) can be analysed by turning on a vacuum expectation value for the adjoint scalar field \(X\) to reduce to the Abelian case. For simplicity we shall focus on the non-invariance of the spatial charges (in other words, defined on an equal-time surface) under spatial deformations, though the extension to the full two-forms discussed above is clear.

\paragraph{D1NC Theory}
In the non-Abelian theory, we can use a (spacetime-dependent) Lie algebra element $\sigma\in\mathfrak{g}$ to define the projection
\begin{equation}
    f^{(\sigma)}_{\mu\nu} = \tr( \sigma F_{\mu\nu} ) \ ,
\end{equation}
of the field strength. The equations of motion of the non-Abelian theory then give us
\begin{subequations}
\begin{align} \label{eq: monopole 'charge'}
    \varepsilon_{ijk} \partial_k f^{(\sigma)}_{ij} = 2 \tr ( D_i \sigma D_i X) \ , \\ \label{eq: electric 'charge'}
    \partial_i f^{(\sigma)}_{0i} = \tr ( D_i \sigma F_{0i} + \mathrm i D_0 X [\sigma, X] ) \ ,
\end{align}
\end{subequations}
as the non-Abelian analogues of the Abelian theory's spatially conserved currents. It is instructive to consider the cases $G = \operatorname{SU}(2)$ and $G = \operatorname{SO}(3),$\footnote{There are two consistent choices of lines for gauge group $\operatorname{SO}(3)$, typically referred to as $\operatorname{SO}(3)_{\pm}$ \cite{Aharony:2013hda}. It is $\operatorname{SO}(3)_+$ that, in the context of 4d MSYM, is related to the $\operatorname{SU}(2)$ theory via S-duality and so we will restrict our attention to this choice of lines here.} taking $\sigma = v^{-1} X$. Using the Prasad--Sommerfield solution \eqref{eq: ps solution} and its higher-monopole cousins, we see any 't Hooft line with integer charge under \eqref{eq: monopole 'charge'} can be screened; this leads to no non-trivial unscreened 't Hooft line for $\operatorname{SU}(2)$, and a single non-trivial unscreened line for $\operatorname{SO}(3)$ with half-integer charge in our normalisation \cite{Aharony:2013hda}. In the former case the $\operatorname U(1)$ magnetic one-form symmetry of the Abelian theory is completely broken, while in the latter it is broken to $\mathbb{Z}_2$. 

The pattern for the electric charge is similar. The computation in section \ref{sec: spectrum} shows that the Wilson lines charged under \eqref{eq: electric 'charge'} can be screened by dyons if they have integer charge: this screens lines for the $\operatorname{SO}(3)$ theory, and leaves the line in the half-integer spin representation of $\operatorname{SU}(2)$. The symmetry breaking is the reverse of before, with a $\mathbb{Z}_2$ electric one-form symmetry group for $\operatorname{SU}(2)$ and none for $\operatorname{SO}(3)$. Of course, this is identical to the symmetry-breaking pattern of the relativistic $\mathfrak{su}(2)$ theories, and so we see that no symmetry enhancement occurs in the low-energy limit. More generally, for an arbitrary group $G$ the same considerations leave a $\operatorname Z(G)$ electric one-form symmetry and a $\operatorname\pi_1(G)$ magnetic one-form symmetry unbroken.

\paragraph{SNC Theory (Galilean Yang--Mills)}

To define the extension of the Abelian spatially conserved quantities to the non-Abelian theory, we take
\begin{subequations}
\begin{align}
    f_{ij}^{(\sigma)} &= \tr (\sigma F_{ij}) \ , \\
    J_i^{(\sigma)} &= \tr (\sigma D_i X) \ ,
\end{align}
\end{subequations}
which then obey
\begin{subequations}
\begin{align}
    \varepsilon_{ijk} \partial_k f_{jk} &= \varepsilon_{ijk} \tr(D_k \sigma F_{ij}) \ , \\ \label{eq: snc electric 'charge'}
    \partial_i J_i^{(\sigma)} &= \tr ( D_i \sigma D_i X + i D_0 X [\sigma, X] - [X,Y^A] [\sigma, Y^A]) \ .
\end{align}
\end{subequations}
In section \ref{sec: spectrum} we constructed solutions for $\mathfrak{g}=\mathfrak{su}(2)$ for which the integrals of both these quantities evaluate to integers when appropriately normalised (note that we must take $\alpha = \delta = 1$ and $\beta = \gamma = 0$ to study the undeformed SNC theory, as we are doing here). As the additional matter contribution in \eqref{eq: snc electric 'charge'} is from an adjoint field, including its charge leaves this conclusion unchanged. We therefore again see that $G= \operatorname{SU}(2)$ has trivial magnetic one-form symmetry and $\mathbb{Z}_2$ electric one-form symmetry, with the reverse true for $\operatorname{SO}(3)$. By the same logic as above this generalises to arbitrary gauge group in precisely the same way. As S-duality maps electrically charged states into magnetically charged states and vice~versa, we see that the duality between the SNC and D1NC theories requires interchanging the gauge group with its dual (as one would expect from relativistic considerations).

\section*{Acknowledgements}
The authors thank Neil Lambert for helpful discussions during the completion of this work and comments on an earlier draft
as well as anonymous referees for helpful suggestions and corrections.

\newcommand\cyrillic[1]{\fontfamily{Domitian-TOsF}\selectfont \foreignlanguage{russian}{#1}}
\newcommand\bulgarian[1]{\fontfamily{Domitian-TOsF}\selectfont \foreignlanguage{bulgarian}{#1}}
\bibliographystyle{unsrturl}
\bibliography{non-lorentzian-theta}

\end{document}